\documentstyle[aps,preprint,floats,epsf,amsmath,amssymb,eqsecnum]{revtex}

\begin{document}

\newcommand{\BR}{{\mathbf{R}}}
\newcommand{\BL}{{\mbox{\boldmath $\lambda$}}}
\newcommand{\BA}{{\mathbf{a}}}

\newcommand{\BRV}{{\langle} {\mathbf{R}}_F^2 (L){\rangle}}
\newcommand{\AVBR} { \overline{{\langle} {\mathbf{R}}_F^2 (L){\rangle}} }

\newcommand{\BRT} {\langle   {\mathbf{R}}_T^2 (L) \rangle }
\newcommand{\ABRT} { \overline{\langle   {\mathbf{R}}_T^2 (L) \rangle }}

\newcommand{\AAT} {\overline{\langle{\mathbf{R}}_T(L) \rangle^2 }}
\newcommand{\BBT} {\langle   {\mathbf{R}}_T (L) \rangle_V^2 }

\newcommand{\BRQ}{{\langle} {\mathbf{R}}_Q^2 (L){\rangle}}
\newcommand{\ABRQ}{ \overline{\langle   {\mathbf{R}}_Q^2 (L) \rangle }}

\newcommand{\APb}{{\overline{\BR}_b}}
\newcommand{\AVR}{{\overline{\BR}}}

\advance\textheight by 0.2in 
\draft

\title{Localization of a polymer in random media:
Relation to the localization of a quantum particle}

\author{Yohannes Shiferaw and Yadin Y. Goldschmidt}

\address{ Department of Physics and Astronomy, \\
  University of Pittsburgh, Pittsburgh PA 15260, U.S.A. }
\date{\today} 

\maketitle

\begin{abstract}
In this paper we consider in detail the connection between the problem
of a polymer in a random medium and that of a quantum particle in a
random potential. We are interested in a system of finite volume where
the polymer is known to be {\it localized } inside a low minimum of
the potential. We show how the end-to-end distance of a polymer which
is free to move can be obtained from the density of states of the
quantum particle using extreme value statistics. We give a physical
interpretation to the recently discovered one-step 
replica-symmetry-breaking solution for the polymer (Phys. Rev. E{\bf
61}, 1729 (2000)) in terms of the statistics of localized tail states.
Numerical solutions of the variational equations for chains of
different length are performed and compared with quenched averages
computed directly by using the eigenfunctions and eigenenergies of the
Schr\"odinger equation for a particle in a one-dimensional random
potential. The quantities investigated are the radius of gyration of a
free gaussian chain, its mean square distance from the origin and the
end-to-end distance of a tethered chain. The probability distribution
for the position of the chain is also investigated. The glassiness of
the system is explained and is estimated from the variance of the
measured quantities.
\end{abstract}
\newpage

\section{Introduction}

  The behavior of polymer chains in random media is a well studied problem
[1-8] that has applications in diverse fields.  Besides the polymers 
themselves this problem is directly related to the statistical mechanics
of a quantum particle in a random potential \cite{gold2}, the
behavior of flux lines in superconductors in the 
presence of columnar defects \cite{nelson,gold3}, and the problem of 
diffusion in a random catalytic environment \cite{nattermann}.
It was found in Ref.~\cite{edwards,cates,nattermann,gold}
that a very long Gaussian chain, immersed in a 
random medium with very short range
correlations of the disorder, will typically curl up in some small region of
low potential energy.  The polymer chain is said to be localized and
for long chains the
end-to-end distance becomes independent of chain length ($R^2\sim L^0$).
For short chains the end-to-end distance scales diffusively ($R^2\sim L$).
In a related paper \cite{gold2}, it was found, using the replica
approach, that a quantum particle exhibits
glassy behavior at low temperatures. The low temperature limit for a 
quantum particle translates into the long chain
limit for polymers. It implies that the free energy landscape
of a long chain is typically very complicated and possesses many 
metastable states.  

Recently a new variational solution has been found for gaussian chain
embedded in a random $\delta$-correlated potential \cite{gold}. This solution,
which involve replica-symmetry-breaking (RSB) at the 1-step level,
gives rise to the correct behavior of the end-to-end distance of the
chain as predicted by the heuristic free-energy estimates of Cates and
Ball \cite{cates}. It predicts the subtle $\log {\cal V}$ dependence
of the chain size, which was not present in the variational solution
of Edwards and Muthukumar \cite{edwards} due to the fact that their solution did
not contain enough variational parameters and hence did not reflect the
translational invariance of the original hamiltonian.

Our aim in this paper is two fold. The first goal is to strengthen and
elucidate the connection between the localization of a polymer in a
random medium and that of a quantum particle. There is an extensive
literature on electron localization that we would like to build
bridges into. The second goal is to give further interpretation and a
better physical picture of the recently discovered variational
solution for a polymer in a random medium \cite{gold}. It turns out
that these two goals intertwine together by using the properties of
the solutions of the Schr\"odinger equation for a quantum particle in a
random environment.

Using the path integral mapping between the partition 
function of a Gaussian polymer chain and the imaginary time 
Schr\"odinger equation, we show 
that the glassy behavior of the polymer chain can be understood
by studying the eigenfunctions of the Schr\"odinger
equation with a random potential.
In particular, the phenomenon of Anderson Localization \cite{anderson,review1}
is crucial to understanding the glassy phase.
We present evidence that shows that polymer localization and 
glassy behavior can 
be traced to the dominance of exponentially localized  tail states.
We also explore various connections between typical glassy behavior, such as 
non-ergodicity and the existence of many metastable states, to the
properties of the eigenfunctions of the Schr\"odinger equation.

In order to describe the glass phase analytically we utilize the variational
replica approach described in \cite{gold} and \cite{gold2}. We further
investigate and interpret the transition discovered there between a replica
symmetric phase to a 1-step replica symmetry breaking phase at a critical
chain length. The 1-step breaking solution describes the localized (glassy)
phase of the polymer and corresponds to the dominance of localized tail
states. In the long chain limit the lowest tail state in each
realization of the disorder will dominate the partition sum. 
On the other hand, the replica symmetric phase corresponds to the
case when a multitude of extended states dominate the partition function. 
Our results are substantiated by a numerical solution of
the problem by solving the imaginary time Schr\"odinger equation
on a lattice in one spatial dimension.

\section{The Model}

The simplest model of a polymer chain in random media is a Gaussian 
(flexible) chain \cite{doi} in a medium of fixed random obstacles
\cite{mut}. In this paper we do not include a self-avoiding interaction.
This model can be described by the Hamiltonian
\begin{equation}
H=\int^L_0 du \left[\frac{M}{2} \left(\frac{d{\mathbf{R}}(u)}{ 
du}\right)^2+\frac{\mu}{2}{\mathbf{R}}^2(u)+V\left({\mathbf{R}}(u)\right)\right],
\label{Ham}
\end{equation}
where ${\mathbf{R}}(u)$ is the $d$ dimensional position vector of a point on the
polymer at arc-length $u\  (0 \leq u \leq L)$, and where $L$ is the 
contour length of the chain (number of links). 
The random medium is described by a random potential $V({\mathbf{R}})$ 
that is taken from a Gaussian distribution that satisfies
\begin{equation}
\langle V({\mathbf{R}}) {\rangle}=0,\;        \; {\langle} 
V({\mathbf{R}})V({\mathbf{R}}'){\rangle}=f\left(({\mathbf{R}}-{\mathbf{R}}')^2\right).
\label{VV}
\end{equation}
The harmonic term in the Hamiltonian is included to mimic the effects of 
finite volume.  This is important to ensure that the model is well
defined, since it turns out that certain equilibrium
properties of the polymer diverge in the infinite volume
limit $(\mu \rightarrow 0)$. The function $f$ characterizes the correlations 
of the random potential, and will depend on the particular problem at
hand. The parameter $M$ is equal to $d/(\beta
b^{2})$, where $\beta =(k_{B}T)^{-1}$, and where $b$ is the Khun bond step.

In this paper we will consider Gaussian distributed random media
defined by the correlation function
\begin{equation}
f\left( (\BR-\BR')^2) \right) = \frac{g}{(\pi \xi^2)^{d/2}}
\exp\left({- ({\mathbf{R-R'}})^2/\xi^2}\right).
\label{Gaussiancor1}
\end{equation} 
Here $g$ determines the strength of the disorder and the parameter 
$\xi$ controls the correlation range of the 
random media.  For $\xi$ large (long range correlations) we expect
to recover the exact results presented in Ref.~\cite{shifgold}.
When $\xi$ is small then we have the more physically relevant case
of short range correlated media. In the limit of $\xi \rightarrow 0$,
$f$ approaches a $d$-dimensional $\delta$-function. This was the case
studied in Refs. \cite{edwards,gold}. 
In the present paper we investigate only the case of short-range 
correlations, i.e. the case of small $\xi$.

Once we have defined the Hamiltonian for any chain configuration ${\mathbf{R}}(u)$,
we can write the partition sum (Green's function) for the set of 
paths of length $L$
that go from ${\mathbf{R}}$ to ${\mathbf{R}}'$ as
\begin{equation}
Z({\mathbf{R}},{\mathbf{R}}';L)=\int_{{\mathbf{R}}(0)
={\mathbf{R}}}^{{\mathbf{R}}(L)={\mathbf{R}}'} 
[d{\mathbf{R}}(u)] {\exp}(-\beta H).
\label{Z}
\end{equation}

All the statistical properties of the polymer will depend on the partition 
sum. For instance, the end-to-end distance (or radius of gyration) of a
polymer chain that is free to move is given by 
\begin{equation}
\overline{{\langle} {\mathbf{R}}_F^2 (L){\rangle}}=\overline{\left(\frac
{\int d{\mathbf{R}} d{\mathbf{R'}}({\mathbf{R}}-{\mathbf{R}'})^2 
Z({\mathbf{R}},{\mathbf{R'}};L)}{\int d{\mathbf{R}} d{\mathbf{R'}}
Z({\mathbf{R}},{\mathbf{R'}};L)}\right)},
\label{rf2av}
\end{equation}  
where the overbar stands for the average of the ratio over the realizations
of the random potential. This average is referred to as a quenched average,
as opposed to an annealed average, where the numerator and denominator are
averaged independently. In some previous studies it has been customary to
replace the quenched average of certain quantities (and in particular
$R_F$) by the more analytically
tractable annealed average. However, this replacement can be justified in
the case of Eq. (\ref{rf2av}) only when the system size is strictly
infinite, since only in that limit can the polymer sample all of space and
find the most favorable potential which will be similar to its environment
in the annealed case. The main problem with this approach is that in
practice we always deal with finite size systems, and in general it is
difficult to assess how big the system size has to be so that the annealed
average is a good approximation to the quenched average. In addition, the
time it takes the chain to sample a large volume is exceedingly long and
unreachable over a reasonable experimental time. In this paper we will
always deal with an explicitly finite system (both analytically and
numerically) and so we will only compute quenched averages. For very large
but finite systems the free chain is said to be {\it localized} in the
sense that its configuration space is dominated by a single
configuration in which the chain is being trapped in a single small
neighborhood. The size of the chain in an uncorrelated random potential 
is given by \cite{cates,gold}
\begin{eqnarray}
R_F \propto (g \ln {\cal{V}})^{-1/(d-4)},
\label{lnV}
\end{eqnarray}
where $\cal{V}$ is the volume of the system. The depth of the well
entrapping the chain is approximately
\begin{eqnarray}
V_{min} \sim (g \ln {\cal{V}})^{2/(d-4)}.
\label{depth}
\end{eqnarray}
This is the binding energy per monomer. Thus the binding energy of the
chain is given by $LV_{min}$.

Another quantity of interest is the averaged mean squared displacement of 
the far end of a polymer with one end that is fixed at the origin.  This is
a measure of the wandering of a tethered polymer immersed in a random medium.
This quantity can be written as 
\begin{equation}
\overline{{\langle} {\mathbf{R}}_T^2 (L){\rangle}}=\overline{\left(\frac{\int 
d{\mathbf{R}} {\mathbf{R}}^2 
Z({\mathbf{0}},{\mathbf{R}};L)}{\int d{\mathbf{R}} Z({\mathbf{0}},{\mathbf{R}};L)}\right)}.
\label{rt2av}
\end{equation}
It is important to distinguish between these two quantities since they
measure different properties of the polymer chain.
Cates and Ball \cite{cates} and Nattermann \cite{nattermann}
discussed the different behaviors of these quantities for a short-range
correlated random potential using heuristic arguments.
In a recent paper \cite{shifgold} we have found analytically that the scaling properties
of $R_F$ and $R_T$ are also very different when the disorder has long
range quadratic correlations. 
This is because in the case of a tethered chain, the free end of the 
chain seeks out favorable regions in the
random medium that are typically very far from the origin.
When both ends are free then the entire chain will simply curl up
in a favorable region in the random medium, and the end-to-end 
distance will not scale as fast. For the tethered chain the quenched
and annealed averages are not expected to coincide even in the
infinite volume limit, since a tethered chain can only sample a finite
volume of size $L^d$, and even for large $L$ the environment near its
tail can not be adjusted since it is immobile.
 
Yet another quantity of interest is 
\begin{equation}
\overline{{\langle} {{\mathbf{R}}^2_Q}(L){\rangle}}=\overline{\left(\frac{
\int d{\mathbf{R}} {\mathbf{R}}^2 Z({\mathbf{R}},{\mathbf{R}};L)}{\int d{\mathbf{R}} 
Z({\mathbf{R}},{\mathbf{R}};L)}\right)},
\label{qdisp}
\end{equation} 
which measures the distance from the origin (of the harmonic
potential) to the average (center of mass) position of a free chain
(more precisely of a free loop).
It is also relevant to the related problem of a
quantum particle in a random potential \cite{gold2,chen}.
The relation between the polymer and the quantum particle problems
arises from the fact that the partition sum of
a polymer chain can be mapped to the density matrix of a quantum particle.
The mapping \cite{gold2,feynman} is given by   
\begin{equation}
\beta \rightarrow 1/\hbar, \; \; L \rightarrow \beta \hbar.
\end{equation}
Then  $\rho(R,R';\beta)=Z(R,R';L=\beta \hbar,\beta=1/\hbar)$ is
the density matrix of a quantum particle at inverse temperature $\beta$.
Note that the variable $u$ is now interpreted as the Trotter (imaginary)
time, and $M$ as the mass of the quantum particle.  Under this
mapping $\overline{{\langle} {{\mathbf{R}}^2_Q}(L){\rangle}}$ can
be interpreted as the average mean squared displacement from the origin 
of a quantum particle in a random plus harmonic potential centered at
the origin. 

Finally we would like to remind the reader the definition of the
density of states for a quantum particle in a random potential. We
first define the quantity $\tilde{\rho}(\beta)$ which is the 
two-sided Laplace transform of the density of states $\rho(E)$:
\begin{eqnarray}
\tilde{\rho}(\beta)=\int_{-\infty}^\infty \exp(-\beta E)\ \rho(E) \ dE.
\label{tilderho}
\end{eqnarray}
This function is given by
\begin{eqnarray}
\tilde{\rho}(\beta)=\lim_{{\cal V}\rightarrow \infty}\frac{1}{\cal V}\int
d\BR \overline{Z(\BR,\BR;\beta)}.
\label{tr2}
\end{eqnarray}
We will make use of these quantities in the following sections.

\section{The Path Integral Mapping}
\label{sec3}

The partition sum of the polymer chain (\ref{Z})
can be mapped to an imaginary time Schr\"odinger equation.
This  mapping (see Ref.~{\cite{feynman}} Eqs.~(3.12)-(3.18)) is given by
\begin{equation}
Z({\mathbf{R}},{\mathbf{R}}';L)=\int_{{\mathbf{R}}(0)={\mathbf{R}}'}^{{\mathbf{R}}(L)={\mathbf{R}}} [d{\mathbf{R}}(u)] {\exp}\left(-\beta H[{\mathbf{R}}(u)]\right)= 
{\langle} {\mathbf{R}}|{\exp}(-\beta L\hat{H})|{\mathbf{R}}'{\rangle},
\label{map}
\end{equation}
where 
\begin{equation}
\hat{H}=-\frac{1}{2M\beta^2}\frac{\partial^2}{\partial
\hat{{\mathbf{R}}}^2}+\frac{\mu}{2}\hat{{\mathbf{R}}}^2 +V(\hat{{\mathbf{R}}}).
\label{hamiltonian}
\end{equation}
So for a given realization of the random potential the polymer
partition sum can be expressed as a matrix element of the
imaginary-time evolution operator.  The matrix elements can
be expanded in a complete set of eigenfunctions of the Hamiltonian operator
to yield
\begin{equation}
{\langle} {\mathbf{R}}|{\exp}(-\beta L\hat{H})|{\mathbf{R}}'{\rangle}=
\sum_{m=0}^\infty {\exp}(-\beta L
E_m) \Phi_m({\mathbf{R}}) \Phi_m^*({\mathbf{R}}'),
\label{eigen}
\end{equation}
where 
\begin{equation}
\hat{H}\Phi_m({\mathbf{R}})=E_m \Phi_m({\mathbf{R}}).
\label{eigenvalue}
\end{equation}
It is clear from the above relation that all properties of the polymer
chain will depend on the eigenvalues and the eigenfunctions of
the Hamiltonian operator $\hat{H}$.  For the problem at hand we will
have to understand these properties  for the case when the potential
$V({\mathbf{R}})$ is random and with the correlations 
given in Eq.~(\ref{Gaussiancor1}) with a small $\xi$.

  The Schr\"odinger equation with a random potential is a 
well known problem that has been intensely studied for a long time
\cite{anderson,review1,frohlich,lifshits}.  
The main property that we will use
is that when $V({\mathbf{R}})$ has short range correlations (i.e. 
correlation length is shorter than any other length scale in
the problem), and if the system size is infinite, then in any
dimension all eigenstates with energy bellow a critical energy $E_M$
( refered to as the mobility edge) are exponentially localized in the form
\begin{equation}
\Phi_m ({\mathbf{R}}) \sim \exp{(- |{\mathbf{R}}-{\mathbf{R}}_m|/\ell_m)}.
\label{localized}
\end{equation}
Here ${\mathbf{R}}_m$ is the localization center of the $m^{th}$
state, and $\ell_m$ is the localization
length of that state. The localization length satisfies 
\begin{eqnarray}
1/\ell_m = \beta\sqrt{2M|E_m|},
\label{ell}
\end{eqnarray}
for $E_m \ll 0$, i.e. deep in the tail region. Intuitively, it is
easy to verify these last two relations for the solution of the Schr\"odinger
equation for a particle in a one dimensional, non-random, attractive 
$\delta$-function potential. This can be thought to represent a local
minimum of the random potential. In $d$-dimensions one can similarly
consider a potential of the form $-v \delta(r)$ and the lowest energy
solution of the radial equation (with zero angular momentum) also
satisfies these relations.
For $E>E_M$ extended states exist when $d>2$. For $d=1,2$ 
there is no mobility edge and all states are exponentially localized.  
The states with energies $E>E_M$ are called extended 
since they are no longer localized but are spread over a finite fraction 
of the system.  Also, it is known that the eigenvalues of the
localized states are discrete, while the eigenvalues
of the extended states form a continuum.  

For finite system size, or if $\mu \neq 0$ in the Hamiltonian given
in Eq.~(\ref{hamiltonian}), the above discussion has
to be modified.  First, the eigenfunctions are always discrete in any
dimensions. But even in one dimension as the energy increases the width
of the localized states eventually becomes comparable to the system size and thus
a localized particle of that energy can go from one end of the sample 
to the other. Thus the distinction between localized and extended
states becomes blurred for a finite system at energies much above the
ground state. Nevertheless, there will still be a  qualitative difference between 
the low energy tail states and the higher energy states with 
large localization lengths. For a finite system (or when $\mu \neq 0$) and for 
any given realization of the random potential there will always be a lowest 
energy state which is by definition the ground state for that realization.
If the volume is large (or $\mu$ small) it will correspond to one of the deep 
tail states of the infinite system.  

In order to study the effect of the eigenfunctions on the physical
properties of the polymer chain, we simply apply Eq.~(\ref{eigen}).
It is clear that when $L$ (or $\beta$) is large the partition sum
is dominated by contributions due to the low energy localized 
states, while if $L$ is small then most of the contributions will
come from the continuum of extended states.  As $L$ is slowly
increased then fewer and fewer localized states contribute
to the sum, until we attain ground state dominance at some
very large $L$.  For simplicity let us concentrate on the
diagonal elements of the evolution operator. Then
\begin{equation}
Z({\mathbf{R}},{\mathbf{R}};L)=\sum_{m} {\exp}(-\beta L
E_m) |\Phi_m({\mathbf{R}})|^2,
\label{loopexp}
\end{equation}  
which is proportional to the probability of finding a loop
of length $L$ passing through the point $\BR$.  Since the
eigenstates for low energies are localized then for large
$L$ we can write 
\begin{equation}
Z({\mathbf{R}},{\mathbf{R}};L) \sim \sum_{m} {\exp}(-\beta L
E_m)  \exp{(-2 |{\mathbf{R}}-{\mathbf{R}}_m|/\ell_m)}.
\end{equation} 
This implies that the probability of finding a long loop passing
through $\BR$ is concentrated around the localization centers
of the low energy eigenstates.  Consequently, a very long 
polymer chain will most likely be found at the ground state
localization center.  When $L$ is small then the partition sum
will be dominated by the extended states, and so the probability
of finding a loop at $\BR$  should be fairly uniform throughout the
system. In a later section we will analyze in detail the evolution of
the partition sum with $L$ by solving the Schr\"odinger equation
on a lattice in $d=1$.

All the physical properties of the polymer chain can be 
expressed in terms of the eigenstates of Schr\"odinger
equation. For instance we can write the end-to-end
distance for a given realization of the random potential
as 
\begin{equation}
\BRV_V=\frac{ 2\sum_{m} \left( a_m \int d\BR \BR^2 \Phi_m^*(\BR)
-\left| \int d\BR \BR \Phi_m(\BR) \right|^2 \right) \exp(-\beta L E_m)
}{\sum_{m} |a_m|^2 \exp(-\beta L E_m)},
\label{exprf}
\end{equation}
where $a_m = \int d\BR \Phi_m(\BR) $, and where 
$\langle \cdot \rangle_V$ refers to a configurational average 
for the case of a fixed realization of random potential.  
When $L$ is large enough so that 
$(E_{1}-E_{gs})L>>1$, where $E_1$ is the eigenvalue of the first
excited state, then only the ground state contributes. In this case we have
\begin{equation}
\BRV_V=2\frac{\int d\BR \BR^2 \Phi_{gs} (\BR)}{\int d\BR \Phi_{gs}(\BR)}
 -2\left( \frac{\int d\BR \BR \Phi_{gs}(\BR)}{\int d\BR \Phi_{gs}(\BR)}
\right)^2 ,
\end{equation}
where $\Phi_{gs}(\BR)$ is the ground state eigenfunction.  It can
be shown that the ground state wave function is positive definite
and so in the large $L$ limit $\BRV^{1/2}$ can be interpreted as the 
width of the ground state eigenfunction.  Assuming the ground
state has the form given in Eq.~(\ref{localized}), we can write
$\BRV_V=2d(d+1)\ell_{gs}^2$, where $\ell_{gs}=\ell_0$ is the 
localization length of the ground state. Upon averaging over all 
realizations of the random potential we get that 
$\AVBR=2d(d+1)\overline{\ell_{gs}^2}$, 
and so the quenched average of the end-to-end distance, in 
the long chain limit, is proportional to the square of the average
localization length of the ground state eigenfunction. We see that
for $\mu \rightarrow 0$ the average is taken over the tail states of 
the Schr\"odinger equation.

When one end of the polymer is tethered to the origin
then the end-to-end distance can be expanded in eigenfunctions
to yield 
\begin{equation}
\BRT_V=\frac{ \sum_{m} \int d\BR \BR^2 \Phi_m^*(\BR) \Phi_m({{\mathbf{0}}}) 
\exp (-\beta L E_m)} {\sum_{m} \int d\BR \Phi_m^*(\BR) \Phi_m({{\mathbf{0}}}) 
\exp (-\beta L E_m)  }.
\label{rtexpand}
\end{equation}
The presence of the term $\Phi_m({{\mathbf{0}}})$ is crucial since 
if the eigenstates are localized then $\Phi_m({{\mathbf{0}}}) \sim \exp
(- |\BR_m|/\ell_m)$, which means that eigenstates localized far
away from the origin may not contribute to the sum even if they
may have very low energies.  
When $L$ is not very large the extended states will dominate the sums in 
Eq.~(\ref{rtexpand}) and the behavior of the end-to-end distance
should be diffusive ($\BRT_V \sim L $). 
When $L$ is large, but before the onset of ground state 
dominance, the sums in Eq.~(\ref{rtexpand}) will be dominated
by localized eigenstates that are centered close to the origin.
The free end of the polymer chain will hop between
localization centers as $L$ increases.  This behavior has been investigated
using Flory arguments in Ref.~\cite{nattermann,zhang} and both 
authors found a
weakly sub-ballistic ( $\ABRT \sim L^2/{\log (L)}^\gamma $ ) behavior 
of the mean squared displacement.  For a finite system and
when $L$ is large enough the ground state will always dominate and we get
$\BRT_V = d(d+1)\ell_{gs}^2 + {\BR^2_{gs}} $.  Since the distance 
from the ground state to the origin will typically be much
larger than the localization length, we find upon averaging
that $\overline{\BRT} \approx \overline{\BR^2_{gs}}$, and so the
quenched average of the end-to-end distance simply converges
to the distance to the localization center of the ground state. This
distance is of the order of the size of the system. Thus the
interesting sub-ballistic $L$-dependence of $\ABRT$ arises
from the contribution of excited states and not from ground state
dominance as for the case of $\AVBR$. 

The mean squared displacement defined in Eq.~(\ref{qdisp})
can also be expanded in eigenfunctions as
\begin{equation}
\BRQ_V=\frac{ \sum_{m} \int d\BR \BR^2 |\Phi_m(\BR)|^2  
\exp (-\beta L E_m)} {\sum_{m} 
\exp (-\beta L E_m)  }.
\end{equation}
In the large $L$ limit we find that 
$\BRQ_V = d(d+1)\ell_{gs}^2/4 + {\BR^2_{gs}} $.  This is
very similar to the case discussed in the previous paragraph
and we find upon averaging
that $\overline{\BRQ} \approx \overline{\BR^2_{gs}}$.
Which implies that $\overline{\BRQ}$ becomes independent of
$L$.  When $L$ is small then the extended states dominate
and we expect a behavior as if there is no random medium.
For a polymer confined by a quadratic potential we find
that $\ABRQ=d (\beta \mu L)^{-1}$.

The relationship between the eigenstates of the Schr\"odinger
equation and the partition sum of the polymer chain
can also be used to understand the behavior of the sample-to-sample variations
of various physical properties of the chain.   The sample-to-sample
variations are important in order to assess whether the average
(over many samples) 
of a physical quantity is an accurate measure of that quantity
for a typical sample.  For instance, the sample to 
sample variation of $\BRV_V$ is defined as 
$\Delta_F=\overline{\BRV^2} - \AVBR^2$.  In the same way we
can define the sample-to-sample variation 
$\Delta_T=\overline{\BRT^2} - \ABRT^2$ and
$\Delta_Q=\overline{\BRQ^2} - \ABRQ^2$.  
So if the sample-to-sample variation of a physical quantity
is large then that implies that the probability distribution
for that quantity is highly skewed and strongly dependent 
on the random sample.  Systems which exhibit such behavior
are said to have glassy properties.
For simplicity if we assume ground state dominance
($L$ very large) then $\Delta_F$ is just a measure of the 
sample-to-sample variation of the localization length of
the ground state.  Also $\Delta_T$ and $\Delta_Q$ are
both a measure of the sample-to-sample variation of the 
distance from the origin to the ground state localization
center.   The crucial point here is that
the sample-to-sample variations are dependent on 
which eigenstates  dominate the partition sum, and so 
there should be a connection between glassy properties and
the behavior of the eigenstates.

\section{The Replica Variational Approach}
\label{sec4}

In this section we review the replica approach that was used
in Refs. \cite{gold} and \cite{gold2}
to compute quenched averages of various physical properties
of the polymer chain. Our goal is to give an interpretation for this
formalism in terms of the localized states picture of the
corresponding Schr\"odinger equation.

In order to compute the quenched average over the random potential we apply 
the replica method.  We first introduce $n$-copies of the system and average 
over the random potential to get
\begin{eqnarray}
Z_n(\{{\mathbf{R}}_a\},\{{\mathbf{R}}'_a\};L)=
\overline{Z({\mathbf{R}}_1,{\mathbf{R}}'_1;L)\cdots
Z({\mathbf{R}}_n,{\mathbf{R}}'_n;L)} \nonumber \\
= \int_{{\mathbf{R}}_a(0)={\mathbf{R}}_a}^{{\mathbf{R}}_a(L)={\mathbf{R}}'_a} 
\prod_{a=1}^n[d{\mathbf{R}}_a]\exp(-\beta H_n),
\label{Zn}
\end{eqnarray}
where
\begin{eqnarray}
H_n=\frac{1}{2}\int_0^L du \sum_{a} \left[ M\left(\frac{d {\mathbf{R}}_a (u)}{d 
u}\right)^2 +\mu {\mathbf{R}}^2_a (u) \right] \nonumber \\
-\frac{\beta}{2} \int^L_0 du 
\int^L_0 du' \sum_{ab} f\left( ({\mathbf{R}}_a (u)-{\mathbf{R}}_b (u'))^2 \right).
\label{Hn}
\end{eqnarray}
The averaged equilibrium properties of the polymer can now be written
in terms of the replicated partition sum ${Z_n}(\{{\mathbf{R}}_a\},\{{\mathbf{R}}_a\};L)$.
For instance, the mean squared end-to-end distance  
defined in Eq.~(\ref{rf2av}) can be 
written in as
\begin{equation}
\overline{{\langle} {\mathbf{R}}_F^2(L){\rangle}}=\lim_{n\rightarrow 0} \,\, 
\frac{\int \prod d{\mathbf{R}}_a
\prod d{\mathbf{R}}'_a \left({\mathbf{R}}_1 - {\mathbf{R}}'_1\right)^2 {Z_n}(\{{\mathbf{R}}_a\},
\{{\mathbf{R}}'_a\};L)}{\int \prod d{\mathbf{R}}_a \prod d{\mathbf{R}'}_a
{Z_n}(\{{\mathbf{R}}_a\},\{{\mathbf{R}}'_a\};L)}.
\label{wanderF}
\end{equation}
Thus, the averaged equilibrium properties of the polymer can be extracted from 
an $n$-body problem by taking the $n\rightarrow 0$ limit at the end. This limit
has to be taken with care, by solving the problem analytically for general $n$,
before taking the limit of $n\rightarrow0$.  Unfortunately the 
replicated partition sum cannot be evaluated 
analytically and a variational approach has been used in Refs.
\cite{gold,gold2} to make further progress. The procedure is to follow
the work of Feynman \cite{feynman} and others \cite{shakhnovich,MP} and
model $H_n$ by a solvable trial Hamiltonian $h_n$ which is determined by 
the stationarity of the variational free energy
\begin{equation}
n \langle F \rangle = {\langle H_n - h_n \rangle}_{h_n} -
\frac{1}{\beta} \ln \int [d\BR_1] \cdots [d\BR_n] \exp(-\beta h_n).
\label{varfree}
\end{equation}
Note that the variational free energy also depends on the boundary 
conditions on the polymer chain.  If we are interested in the
case when one end is fixed then the partition sum should
be over paths with one end fixed.  In this case the path
integrals should be evaluated using 
$\int d\BR_i \int^{\BR_i(L)=\BR_i}_{\BR_i(0)={\mathbf{0}}} [d\BR_i]$.
When both ends are free one should use instead
$\int d\BR_i d\BR_i'  \int^{\BR_i(L)=\BR_i}_{\BR_i(0)={\BR_i'}} [d\BR_i]$.
Alternatively,  one can evaluate the partition sum with
$\int d\BR_i \int^{\BR_i(L)=\BR_i}_{\BR_i(0)=\BR_i  } [d\BR_i]$,
which yields the free energy of a polymer loop of contour length $L$.  
The parameters that minimize the variational free energy for this case should
be the same as for the case when both ends are free since in both
cases the polymer loop can be anywhere in the system. These are 
the boundary conditions
that were used in Ref. \cite{gold}, where only the polymer chain that
is free to move was investigated.  

The quadratic trial Hamiltonian has been parametrized \cite{gold} by
\begin{eqnarray}
h_n=\frac{1}{2}\int_0^L du \sum_{a} \left[ M\left(\frac{d {\mathbf{R}}_a (u)}{d 
u}\right)^2 +\mu {\mathbf{R}}^2_a (u) \right] \nonumber \\
-\frac{1}{4L}\int^L_0 du 
\int^L_0 du' \sum_{ab} p_{ab} \left( \BR_a (u) - \BR_b (u') \right)^2,
\label{trialh}
\end{eqnarray}
where the matrix elements $p_{ab}$ are the variational
parameters.  The physical motivation for this ansatz is that
the replica-replica interaction in the original Hamiltonian
is modeled by a quadratic interaction which can be different
for different replica pairs.  Also, as noticed in \cite{gold},
the quadratic interaction has the same translational invariance
as the interaction term in the original Hamiltonian.   
The case when the replica-replica
interactions are the same ($p_{ab}=constant$) corresponds to the
case of long range quadratic correlations which was solved exactly in
\cite{shifgold}.  If we expand and simplify the quadratic 
interaction we can rewrite the trial Hamiltonian as
\begin{eqnarray}
h_n=\frac{1}{2}\int_0^L du \sum_{a} \left[ M\left(\frac{d {\mathbf{R}}_a (u)}{d 
u}\right)^2 +\lambda {\mathbf{R}}^2_a (u) \right] \nonumber \\
+\frac{1}{2L}\int^L_0 du 
\int^L_0 du' \sum_{ab} p_{ab} \BR_a (u) \cdot \BR_b (u'),
\label{trialh2}
\end{eqnarray}
where $\lambda=\mu - \sum_b p_{ab}$. Here, $\lambda$ is assumed to be
independent of the replica index $a$, as is the case if $p$ is a
hierarchical matrix. In Ref. \cite{gold} $\lambda$ was treated
as an independent
variational parameter, and the condition $\lambda=\mu-\sum_b p_{ab}$
emerged automatically as a result of the translational invariance of
the disorder dependent term in the original hamiltonian $H_n$.

In the 1-step RSB scheme, the matrix $p_{ab}$ can be parametrized as
$(\tilde{p},p(x))$ with 
\begin{eqnarray}
p(x)= \left\{ \begin{array}{l}
-s_0 \ \ \ \ \ 0<x<x_c \  \\
-s_1 \ \ \ \ \ x_c<x<1 \ ,
\end{array} \right.
\label{pmat}
\end{eqnarray}
and where $x$ is Parisi's replica index. In \cite{gold} $\tilde{p}$ has been denoted by
$\lambda-\lambda_1$. Thus, five variational parameters were used: $\lambda$,
$\lambda_1$, $s_0$, $s_1$ and $x_c$. The variational free energy was
expressed as a function of these variational parameters. Taking the
partial derivative with respect to the variational parameters and
equating them to zero one gets five non-linear equations which could be solved
analytically when $L$ was large and $\mu$ was small \cite{gold}. In this paper we
started from the same free energy and took its partial derivatives
without simplifying the expressions for large $L$. As mentioned above 
because of the consequences of translational invariance we could reduce the
number of parameters and equations to four, since
\begin{eqnarray}
\lambda_1+x_c s_0+(1-x_c)s_1=\mu \ .
\label{trans}
\end{eqnarray}
We proceeded to solve the equations of stationarity numerically using 
a standard iterative method \cite{press}.  
Once the variational parameters have been found we can obtain
expressions for various physical quantities.  For instance,
we find that
\begin{equation}
\overline{{\langle} {\mathbf{R}}_F^2 
(L){\rangle}}=\frac{2d}{\beta\sqrt{M\lambda}}
\frac{\sinh\left(\sqrt{\frac{\lambda}{M}}L\right)}
{\left(\cosh\left(\sqrt{\frac{\lambda}{M}}L\right)+1\right)},
\label{rsqf}
\end{equation}
and also that
\begin{equation}
\overline{{\langle} {\mathbf{R}}^2_Q (L){\rangle}}
=\frac{d}{\beta L} \left(
\frac{1}{\mu}-\frac{1}{\lambda} +
\frac{s_0}{\mu^2}+\frac{(1-x_c)\Sigma}{x_c \mu (\mu+\Sigma)} 
\right)
+\frac{d}{2\beta\sqrt{M\lambda}}
\coth\left(\sqrt{\frac{\lambda}{M}}
\frac{L}{2}\right),
\label{shift}
\end{equation}
where we have put $\Sigma=x_c (s_1-s_0)$.
Details of the numerical results will be given in the next section.
For large $L$ and small $\mu$ it can be seen using the results obtained
in \cite{gold} that to leading order
\begin{equation}
\overline{{\langle} {\mathbf{R}}_F^2 
(L){\rangle}}=\frac{2d}{\beta\sqrt{M\lambda}},
\end{equation}
and
\begin{equation}
\overline{{\langle} {\mathbf{R}}^2_Q (L){\rangle}}
=\frac{1}{\beta \mu L x_c},
\end{equation}
with
\begin{equation}
\lambda=\frac{d^{4/(4-d)}}{(2\pi)^{2d/(4-d)}}\left(  \beta^{2}M\right)
^{(4+d)/(4-d)}\left(  g\ \left|  \ln\mu\right|  \right)  ^{4/(4-d)},%
\label{lamfin}%
\end{equation}
and
\begin{eqnarray}
x_{c}=\frac{1}{L}\left(  \frac{d^{d-2}}{(2\pi)^{d}}g^{2}\beta^{d+4}%
M^{d}\left|  \ln\mu\right|  ^{d-2}\right)  ^{-1/(4-d)}.
\label{xcf}%
\end{eqnarray}

\section{Numerical Procedure}
\label{s5}

We check the validity of the analytic solution by numerically
computing the quenched average of various physical properties
of the polymer. This is computationally intensive 
because all quantities will have to be averaged over many
realizations of the random potential.  In this paper we 
will only concentrate on the case
$d=1$.  Although this does not correspond to a physical polymer ($d=3$)
we will still be able to check  the validity of our analytical results for 
the special case $d=1$.   

We evaluate numerically the right hand side of Eq.~(\ref{eigen}) by solving 
the Schr\"odinger equation on a one dimensional lattice of $N$
sites \cite{press}.  The lattice Hamiltonian is an $N\times N$ matrix
with matrix elements given by 
\begin{equation}
H_{ij}=-\frac{1}{2M\beta^2\Delta^2} \left( \delta_{i,j+1}+\delta_{i+1,j} 
\right)  +
\left(\frac{\mu}{2}\Delta^2 (i-N/2)^2 + V(i) \right) \delta_{i,j}
\end{equation}
where the lattice spacing is $\Delta=S/N$, and where $S$ is the system
size.  Since we are interested in the continuum limit  $\Delta$ will be
kept small.  Note that the index $i$ corresponds to the 
position $R_i=\Delta i$. We impose hard wall boundary conditions at
the end of the lattice. The eigenvalues and eigenvectors can 
now be found directly by diagonalizing the matrix using a standard numerical
routine \cite{press}.  Once these are known we can construct the
partition sum at any value of $L$ using Eq.~(\ref{eigen}).  
The partition sum can then be used to compute the quantity
of interest, such as $\BRV_V$.  We repeat this procedure a large
number of times and average the results to get a numerical
approximation to $\AVBR$.

The correlated Gaussian random potential described by Eq.~(\ref{Gaussiancor1})
is modeled by a sequence of 
$N$ numbers $\{ V_{\xi}(i) \}_{i=1,..,N}$ that 
obey $\langle V_{\xi}(i)V_{\xi}(i+l) \rangle \propto 
{\exp}\left(- \Delta^2 l^2/\xi^2 \right)$.
These numbers are then placed on a lattice of $N$ sites
in the given order.  To generate such numbers we use an
established  method for generating correlated random
numbers.  The details of this method are
described in Ref.~\cite{shifgold}.

\section{Results and Discussion}
\subsection{Numerical and Analytical results for $\AVBR$ and $\ABRQ$}

\begin{figure}
\centerline{\epsfysize 8.5cm \epsfbox{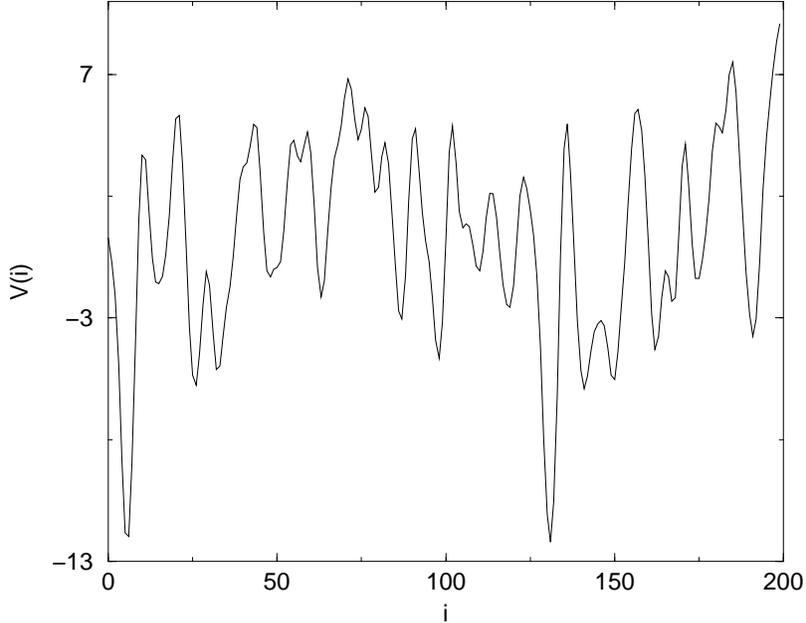}} \vspace{3mm}
\caption{A typical random sample with $\xi=1/\sqrt{2}$}  
\label{sample1}
\end{figure}

Using the method described in the previous section we 
compute $\AVBR$ on a lattice of size $N=200$ (S=40). 
Here, we concentrate on the case where the random potential has 
very short range correlations.  We generate random
potential samples with correlation length  $\xi=1/\sqrt{2}$.
In Fig.~\ref{sample1} we show a typical sample of
$200$ numbers from such a distribution.    
Notice that the short range correlation leads to a rugged random potential
landscape.
The average mean squared displacement $\AVBR$
is computed by averaging over $10000$ samples, and the error
is estimated by computing the standard deviation of $10$ sets
of $1000$ samples.

We were able to find numerical solutions to the 
non-linear stationarity
equations.  We found that for a given set of parameters
there is a chain length $L_c$ (which depends on the strength of the
disorder) such that for $0<L<L_c$
there is only a replica symmetric solution.  This is the case
when the variational parameters satisfy $x_c=1$ and $s_0=s_1$.
For $L>L_c$ there is still a replica symmetric solution 
but we also find an additional replica symmetry breaking
solution. So in this regime we find an additional solution such that
$0<x_c< 1$ and $s_0 \neq s_1$.  In order to decide which solution
correctly describes the physics in that regime  we compare their 
respective predictions
to the lattice computation of $\AVBR$ and $\ABRQ$.

\begin{figure}
\centerline{\epsfysize 8.5cm \epsfbox{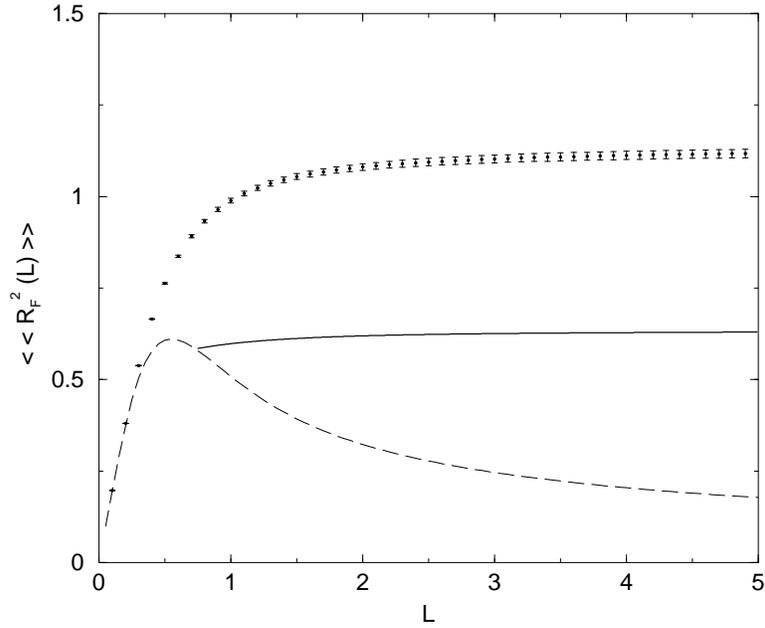}} \vspace{3mm}
\caption{Plot of $\AVBR$ vs. $L$.  The parameters are 
$M=1/2$, $g=25$, $\beta=1$, $\xi=1/\sqrt{2}$, and
$\mu=0.01$.  The dotted line is generated by averaging over 
$10000$ samples on a lattice of size S=40 with $\Delta=0.2$. The error bars
are found by computing the standard deviation of $10$ sets
of $1000$ samples.  The dashed line 
is the RS solution, and the solid line is the RSB solution.
}  
\label{rf1}
\end{figure}

In Fig.~\ref{rf1} we plot the mean squared displacement
$\AVBR$ vs. $L$ for a given set of parameters.
We plot this quantity using the lattice result,
and also using the two predictions of the variational method.
Note that in the labels of the plots the average over the disorder is
denoted by a second set of brackets rather than an overbar.
For $L$ below $L_c \approx 0.73$ there is only a RS solution
which is very close
to the lattice prediction.  For $L$ greater than $L_c$ the 
RS and RSB solutions are different and it is clear that the
RSB solution is closer to the lattice result.
We can see that the end-to-end distance saturates
at a constant value as $L$ increases.  This behavior is correctly
predicted by the RSB solution but not by the RS solution.

\begin{figure}
\centerline{\epsfysize 8.5cm \epsfbox{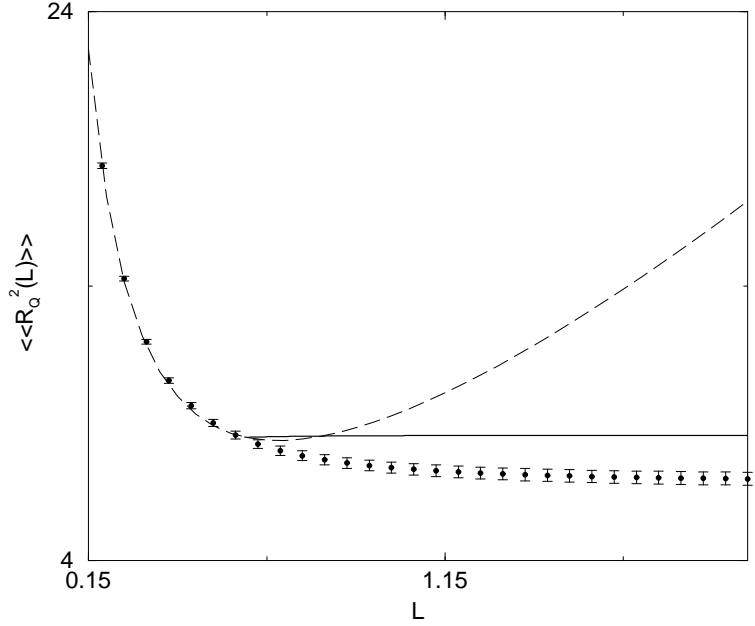}} \vspace{3mm}
\caption{Plot of $\ABRQ$ vs. $L$.  The parameters are the same
as those of Fig.~\ref{rf1} except that here we use $\mu=0.3$.  
The dotted line is generated by averaging over 
$10000$ samples. The error bars 
are found by computing the standard deviation of $10$ sets 
of $1000$ samples. The dashed line 
is the RS solution, and the solid line is the RSB solution.}  
\label{rq1}
\end{figure}

We now turn our attention to the quantity $\ABRQ$.  In Fig.~\ref{rq1}
we plot $\ABRQ $ vs. $L$ using the lattice computation and also
using the RS and RSB solutions.  We can clearly see that when
the RS solution differs from the RSB solution, the 
RSB prediction is closer to the lattice prediction.  Again,
the quantity $\ABRQ$ becomes constant for large $L$ and this is
correctly predicted by the RSB solution. For this quantity similar
results were obtained previously in Refs. \cite{gold2} and \cite{chen}. 

It is clear from
Fig.~\ref{rf1} and \ref{rq1} that the variational method with
the quadratic ansatz in Eq.~(\ref{trialh}) is quite effective in describing
the physical properties of the polymer chain.  The features
predicted by the lattice computation are consistent with a
RS solution for chain lengths shorter than $L_c$ and a RSB solution for
chains longer than $L_c$. In a later section we
will explore the physical interpretation of the variational solution
and show that it is indeed consistent with the physics of the
problem.

\subsection{Localized eigenstates and glassy behavior}

\begin{figure}
\centerline{\epsfysize 8.5cm \epsfbox{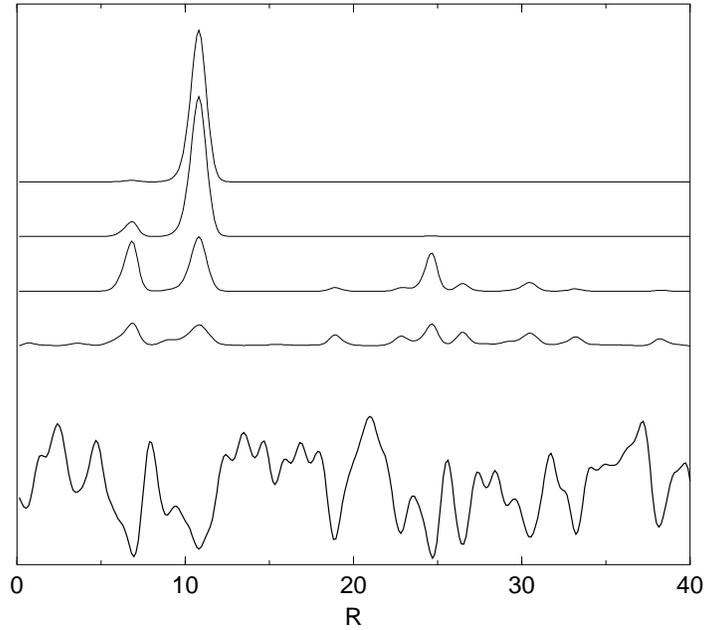}} \vspace{3mm}
\caption{Plot of $P(R,L)$ vs $R$ for four values of $L$. The bottom
most curve is the random potential sample that is used. From bottom
to top we use $L=.3,1,10,20$. The parameters are the same as those
used for Fig.~\ref{rf1}.}  
\label{dist}
\end{figure}

In this section we explore, using the lattice computation,
the connection between the eigenstates of the Schr\"odinger equation and
the physical properties of the polymer chain.  We focus
on the probability distribution defined as $P(\BR,L)=Z(\BR,\BR,L)/
\int Z(\BR,\BR,L) d\BR$ which can be interpreted as the probability
of finding a closed polymer chain of length $L$ which passes through
the point $\BR$ (for a given realization of the random potential).  
We consider this probability distribution since it 
gives the most direct connection
between the chain properties and the eigenfunctions of the 
Schr\"odinger equation.  In Fig.~\ref{dist}
we plot $P(R,L)$ vs. $R$ for four different chain lengths.
We also include a  plot of the random potential sample that is used.
We can see clearly that the probability distribution evolves
from a flat distribution to distributions that are sharply peaked at various
locations in the sample.  Also, an important feature is that the
number of peaks decreases with increasing $L$ until finally
there is only one peak.  This implies that longer chains
have a tendency to be found in a few favorable regions in the
sample, while short chains can be found with equal probability
almost anywhere in the sample. 
Another observation is that the peaks in the distribution, for
the cases $L=1,10,20$, are concentrated
around the valleys of the random landscape.  This shows that the 
chain is more likely to be found in regions of low average potential.
As can be seen the width of the well is also important.

\begin{figure}
\centerline{\epsfysize 8.5cm \epsfbox{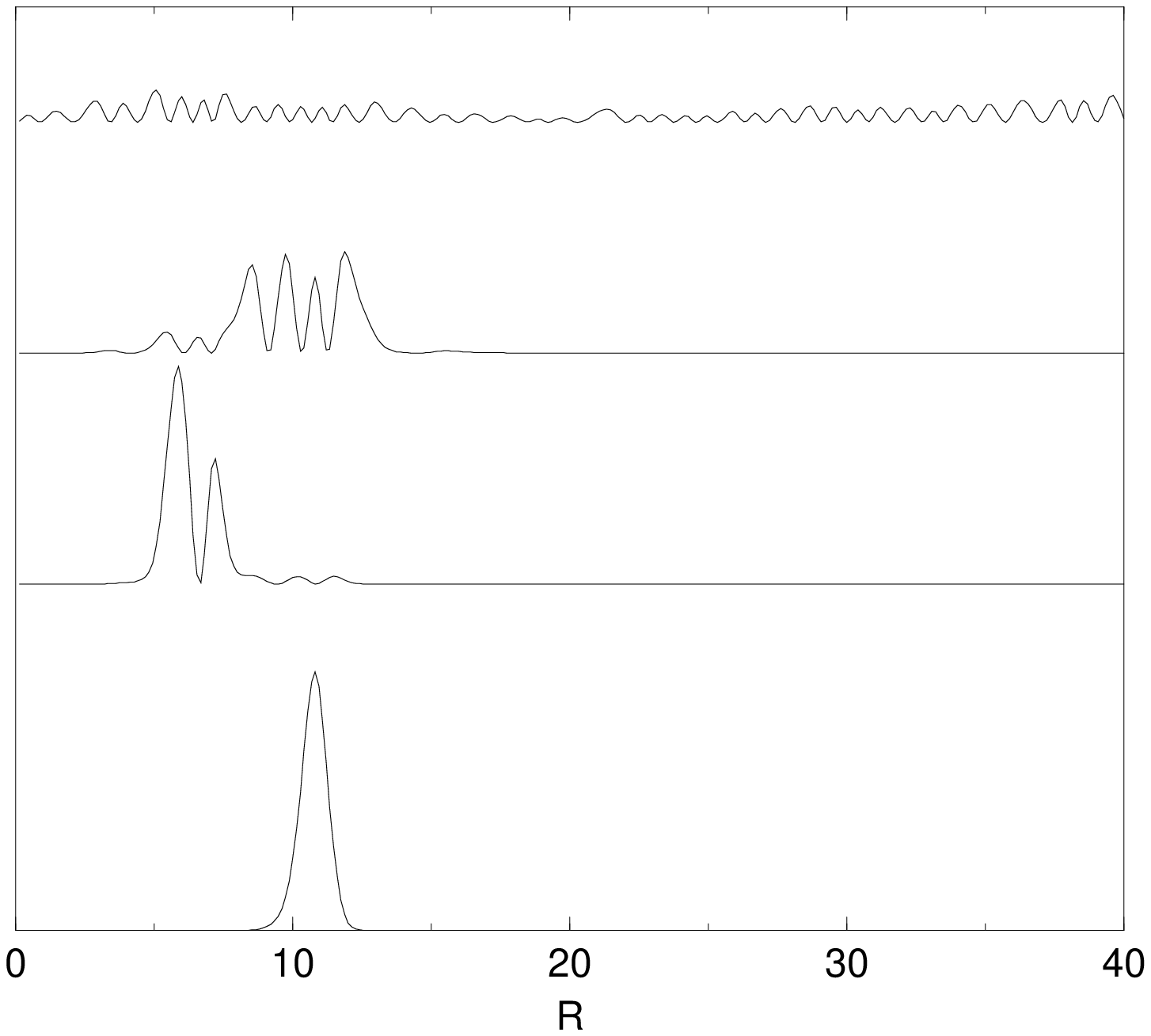}} \vspace{3mm}
\caption{Plot of $\Phi^2_m(R)$ vs. $R$. 
From bottom up we plot
the eigenfunctions with $m=0,10,19,39$. We use a lattice
of size S=40 with $\Delta=0.133$. All other parameters are the 
same as that of Fig.~\ref{rf1}.}  
\label{eig}
\end{figure}

\begin{figure}
\centerline{\epsfysize 8.5cm \epsfbox{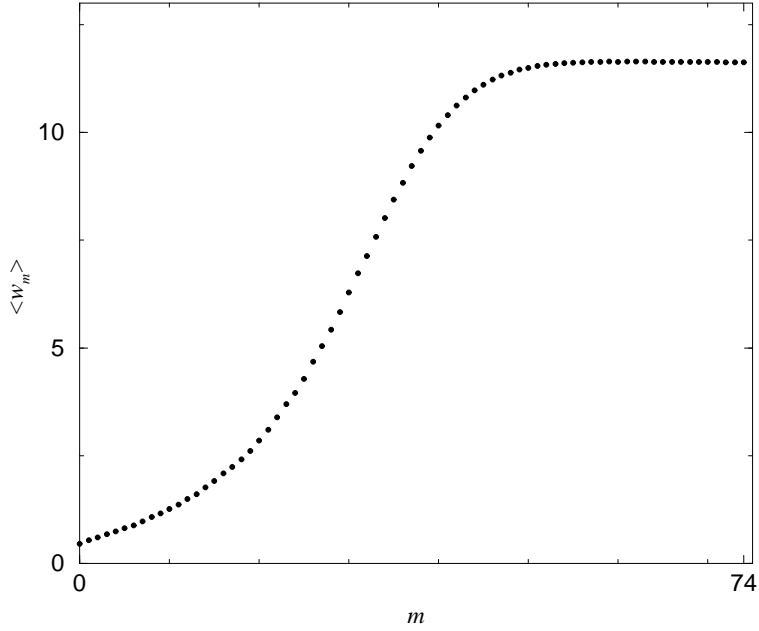}} \vspace{3mm}
\caption{Plot of $\overline{w}_m$ vs. $m$. We use a lattice
of size S=40 with $\Delta=0.133$ and plot $\overline{w}_m$ for the first $75$
eigenstates. We use $10000$ samples and the parameters are the same 
as that of Fig.~\ref{rf1}. }  
\label{width}
\end{figure}

\begin{figure}
\centerline{\epsfysize 8.5cm \epsfbox{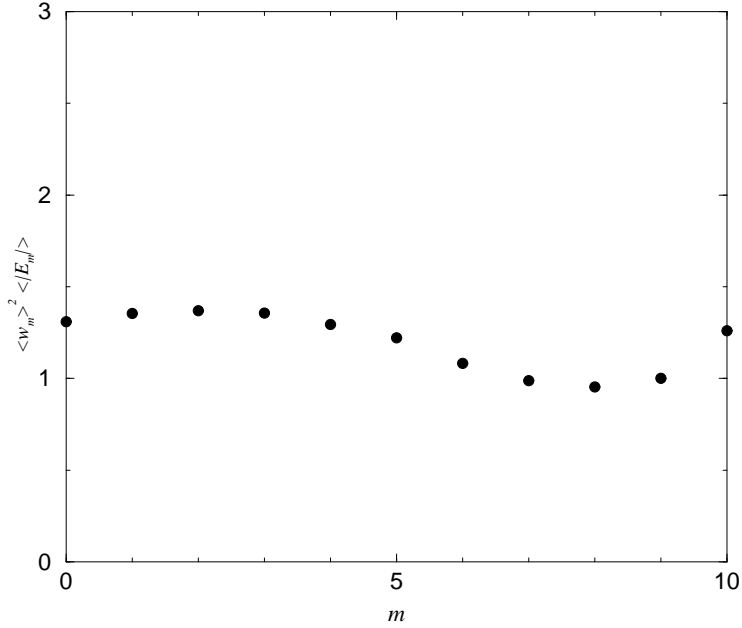}} \vspace{3mm}
\caption{Plot of  $\overline{w}_m^2 |\overline{E}_m|$
 vs. $m$ for the lowest ten eigenfunctions. We average over
$10000$ samples and the parameters are the same as that of Fig.~\ref{rf1}. }  
\label{wvse}
\end{figure}

The results of the previous section can be better understood
by studying the properties of the eigenfunctions of the Schr\"odinger
equation in a random potential.  Using the eigenfunction
expansion in Eq.~(\ref{loopexp}) we can see that $P(\BR,L)$ is just
a sum of the eigenfunctions squared weighted with the Boltzman factor
$e^{-\beta LE_m}$.  So the shape of $P(\BR,L)$  is 
dependent on which eigenstates
have the dominant weight at a given chain length $L$. 
In Fig.~\ref{eig} we plot a representative sample of the 
eigenfunctions for a typical random potential sample.  
In order to get a more general picture of the
structure of the eigenfunctions we measure for each eigenfunction
the width defined by 
\begin{equation}
w_m^2= \int R^2 \Phi^2_m(R) dR- \left( \int R \Phi^2_m(R) dR\right)^2.
\label{widtheq}
\end{equation}
This width is closely related to the localization length $\ell_m$
defined earlier. 
We then average this quantity over many realizations of the
random potential and in
Fig.~\ref{width} we plot $\overline{w}_m$ vs. $m$.
It is clear that as $m$ is increased the width of the eigenfunctions
also increase.  Clearly, on average the ground state is the state with the
smallest width.  This explains
why the the distribution $P(\BR,L)$ evolves with $L$ into
fewer and fewer sharply localized peaks.  This is just because
the low energy eigenfunctions are sharply localized and they dominate
the partition sum as $L$ is increased.  The higher excited states
have typically large localization lengths, and these are the states
that dominate the partition sum when $L$ is short. For the lowest
states we have verified the relation $w_m^2 \sim 1/|E_m|$ . In
Fig.~\ref{wvse} we plot $\overline{w}_m^2 |\overline{E}_m|$
vs. $m$ for the lowest ten eigenfunctions and show that
it is nearly constant in agreement with Eq.~(\ref{ell}). 

\begin{figure}
\centerline{\epsfysize 8.5cm \epsfbox{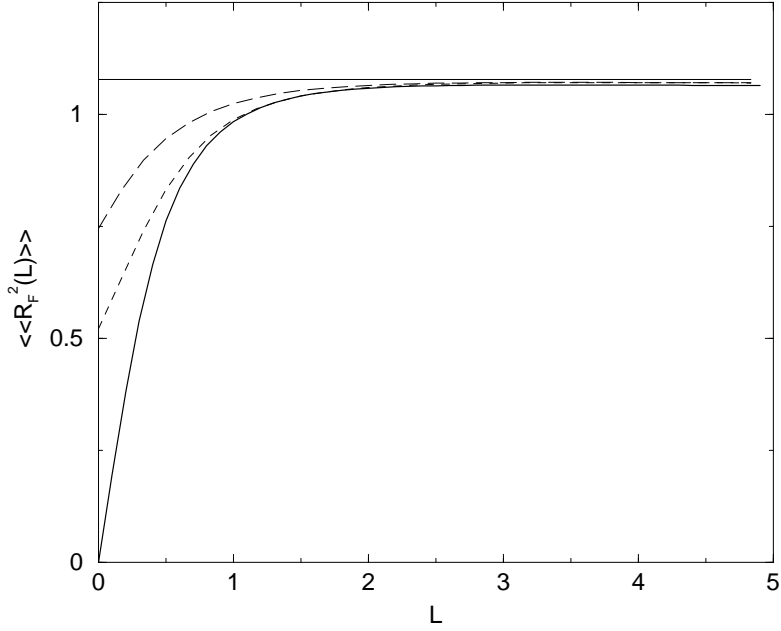}} \vspace{3mm}
\caption{Plot of $\AVBR$ vs. $L$. The thick solid line is computed using
all the eigenfunctions. The thin solid line is found using only the
ground state.  The long dashed line is using the first five
eigenfunctions and the short dashed line is using the first
ten eigenfunctions.}  
\label{stt}
\end{figure}

We now consider the role of the eigenfunctions on 
averaged quantities such as $\AVBR$ and $\ABRQ$.
In order to study the effect of the eigenfunctions
on $\AVBR$ we truncate the sum over eigenstates 
$\sum_{0}^{N}$ to $\sum_{0}^{j}$
in the eigenfunction expansion in Eq.~(\ref{exprf}).  
So by changing the index $j$
we can see how the average end-to-end distance depends on how many 
eigenfunctions are kept in the expansion.  In Fig.~\ref{stt} we plot
the average end-to-end distance $\AVBR$ 
for a number of different $j$ values.  As
expected the flat portion of the graph when $L$ is large is entirely due to
the ground state contribution.  So the constant $\AVBR$ is just a 
consequence of
ground state dominance.  This is consistent with the 
results in Sec.~\ref{sec3} where we showed that 
for long chains $\AVBR=2d(d+1)\overline{(\ell_{gs}^2)}$
which is indeed independent of $L$.
We can also see that the curved portion
of the curve corresponds to the case before ground state
dominance when a number of tail states contribute to the partition
sum.  For small $L$ we see that more than ten eigenfunctions 
are needed in order to capture the correct behavior of $\AVBR$.   

\begin{figure}
\centerline{\epsfysize 8.5cm \epsfbox{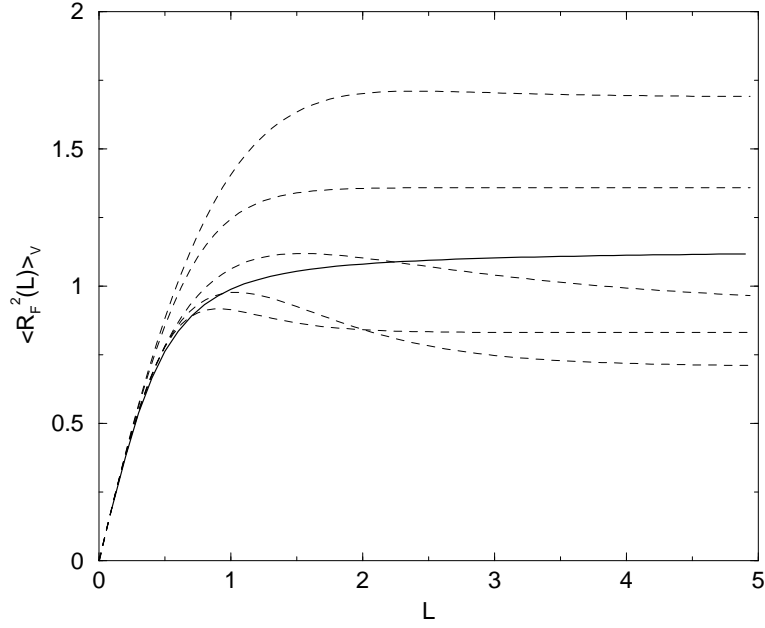}} \vspace{3mm}
\caption{Plot of $\BRV_V$ vs. $L$. The dashed lines correspond
to five randomly chosen samples.  The thick line is the 
average $\AVBR$ over $10000$ samples.
\label{rfa}
}
\end{figure}

\begin{figure}
\centerline{\epsfysize 8.5cm \epsfbox{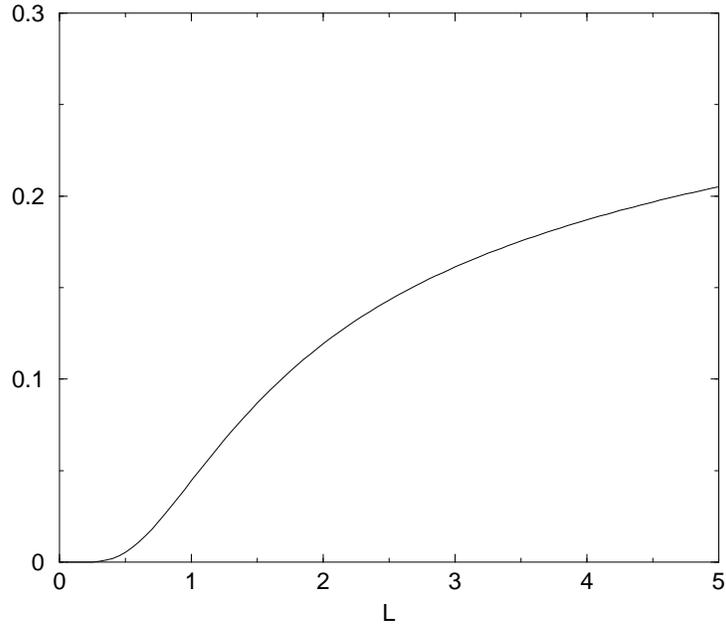}} \vspace{3mm}
\caption{Plot of $\Delta_F/\AVBR$ vs. $L$. The parameters are the
same as in Fig.~\ref{rf1} and we average over $1000$ samples.
\label{delf}
}
\end{figure}

As discussed in the previous paragraphs it is clear that the evolution
of the partition sum is dependent on the nature of the eigenfunctions.
Consequently, the sample-to-sample variation of quantities such
as $\BRV_V$ and $\BRQ_V$ are also crucially dependent on the 
eigenfunctions.  These sample-to-sample variations are important
since they provide a measure of the glassiness of the system.
In Fig.~\ref{rfa} we plot $\BRV_V$ vs. $L$ for $5$ randomly 
picked samples. 
Notice that the spread of $\BRV_V$ around the average increases with $L$.
In order to quantify this spread more accurately we plot the relative
sample-to-sample fluctuations
$\Delta_F/\AVBR$ vs. $L$ in Fig.~\ref{delf}. 
It is clear again that as the
the length of the chain is increased the sample-to-sample
fluctuations also increase.  Comparing Fig.~\ref{delf}
with Fig.~\ref{stt} it is clear that the sample
to sample fluctuations increase rapidly at the same approximate value
of $L$ at which the low energy eigenstates begin to dominate
the behavior of $\AVBR$.  The reason for this is that
distributions dominated
by localized states are strongly sample dependent.
On the other hand when $L$ is small and the high excited
states dominate, the sample-to-sample fluctuations are small
since the shape of the extended states do not depend strongly
on the random sample.
So we can conclude that for a polymer in short range correlated
random media, the glassy characteristics are directly related to the 
dominance of localized eigenfunctions.

\subsection{Analysis of $\ABRT$}

We will now consider the end-to-end distance $\ABRT$.
Since one end is tethered to the origin
it turns out that this quantity behaves very differently
from the quantities $\AVBR$ and $\ABRQ$.
The replica variational approach described in Sec.~\ref{sec4}
is not adequate in this case in this case and hence we
will only present the results of the lattice computation. 

For short range correlations we found that the numerical
method described in Sec.~\ref{s5} was unreliable.  The reason for this
is that the sum over energy eigenfunctions in Eq.~(\ref{eigen}) is unstable 
since the overlap $\Phi_m (R)\Phi_m (R')$ (for
short range correlations) is a number
on the order of $\exp(-|R-R'|/\ell_m)$ which is typically extremely
small.  However, we were able evaluate Eq.~(\ref{eigen}) 
accurately for small $\xi$ by solving
the Schr\"odinger equation on a lattice using  a fourth
order Runge-Kutta algorithm with a very small time step ($t \sim
10^{-3}$).  We performed the computation on a large system
($S=160, N=800$) in order to minimize finite size effects.

\begin{figure}
\centerline{\epsfysize 8.5cm \epsfbox{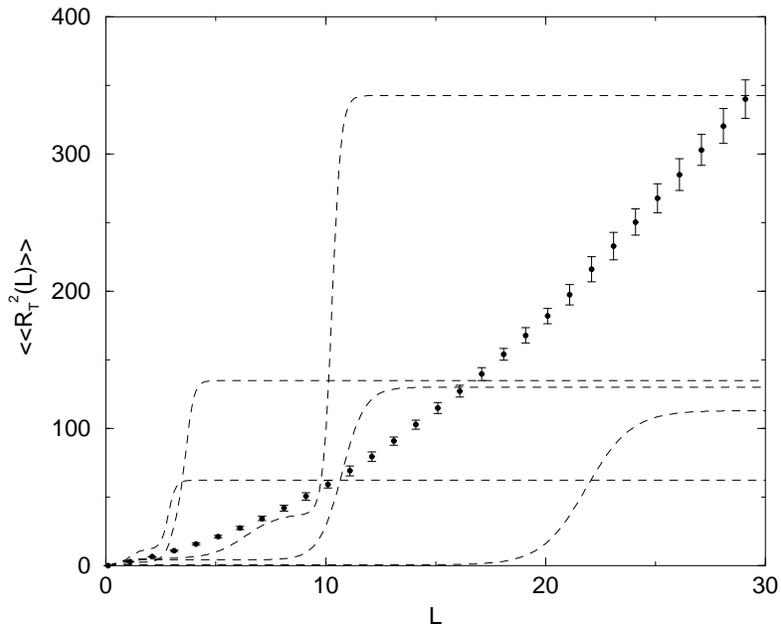}} \vspace{3mm}
\caption{Plot of $\ABRT$ vs. $L$. The full circles are generated by
averaging over $10000$ samples, and the error bars are found
by computing the standard deviation of $10$ sets of $1000$ samples.  
The dashed lines are plots of $\BBT$ vs. $L$ for five
typical samples.  The parameters are the
same as in Fig.~\ref{rf1}, only that here the system 
size is four times larger ($S=160$) and we set $\mu=0$.
We use a lattice of $800$ sites with $\Delta=0.2$.}  
\label{rt1}
\end{figure}

In Fig.~\ref{rt1} we plot $\ABRT$ vs. $L$
and also include $\BBT$ vs. $L$ for five typical samples.
Notice that for a given sample the end-to-end distance
$\BBT$ has a tendency to change rapidly and then remain constant.
As discussed in Sec.~\ref{sec3} this is simply a consequence
of hopping between localized states.  That is, the free end
of the polymer rapidly finds a deep well (which is also a localization
center) and stays there. Upon averaging over $10000$ random samples
we find that  $\ABRT$ grows 
linearly with $L$ for small $L$ but scales faster than diffusion
for large $L$.  The diffusive behavior when $L$ is small is 
not surprising since a short chain should not be 
affected by the random media. 
The claim based on Flory type arguments 
\cite{nattermann,zhang} is that  $\ABRT \sim L^2/(\log(L))^\gamma$
when $L$ is large.  However, we were not able to extract 
a consistent exponent $\gamma$ for the range of $L$ that we 
explored.  The reason for this is that our data is not accurate
enough to detect the precise logarithmic 
correction to scaling.  However we went ahead and 
tried to find the best power law fit $\ABRT \propto L^{\nu}$ 
for large $L$.  We found that for $5 \lesssim L \lesssim 25 $ the exponent
$\nu \approx 1.74$ yielded an excellent fit to the data.
This result is consistent with the sub-ballistic prediction 
as it is very close but slightly less than ballistic scaling ($L^2$).

\subsection{Analysis of the variational solution}

\begin{figure}
\centerline{\epsfysize 8.5cm \epsfbox{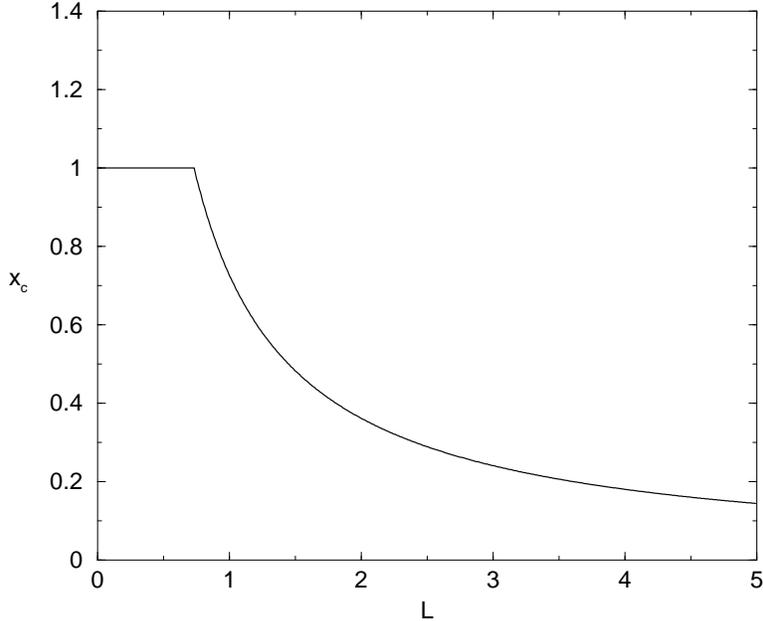}} \vspace{3mm}
\caption{Plot of $x_c$ vs. $L$. The parameters are the same 
as those used in Fig.~\ref{rf1}.
\label{zplot}
}
\end{figure}

So far we have seen that the glassy characteristics of
a polymer in a short range correlated random potential
is closely related to the dominance of low energy
eigenfunctions.  We also know that the variational
solution possesses a RS solution for $L<L_c$ and an
RSB solution for $L>L_c$. It is well known that 
replica symmetry breaking is typically associated with 
glassy behavior, and for our model we show that the onset
of RSB is precisely when the system begins to exhibit
glassy behavior.  The variational parameter that best 
reveals the transition between RS and RSB is the 
break point $x_c$.  If $x_c=1$ then there is only an
RS solution, and if $0<x_c<1$ then that corresponds to
an RSB solution.  In Fig.~\ref{zplot} we plot $x_c$ vs. $L$
using the same parameters that were used in Fig.~\ref{rf1}.
We can see that onset of the RSB solution is at $L_c \approx .73$.
If we compare this result to the plot of $\Delta_F/\AVBR$ vs $L$
in Fig.~\ref{delf},
we see that near $L_c \approx 0.5$ the sample-to-sample fluctuations
begin to rise rapidly.  This result provides strong evidence
that when the RSB solution is valid the polymer chain does indeed
exhibit glassy behavior.

In Ref.~\cite{gold} approximate analytic solutions to the
variational equations were found.  This was for the case of 
large $L$, small $\mu$, and a delta correlated random potential
($\xi \rightarrow 0$).  It was found that
\begin{equation}
x_c=\frac{1}{L}\left(  
\frac{d^{d-2}}{(2\pi)^d}g^2 \beta^{d+4} M^d |\log \mu|^{d-2}
\right)^{-1/(4-d)}.
\label{zequation}
\end{equation}
We checked the $L$ dependence of $x_c$ in Fig.~\ref{zplot} and
indeed we found that $x_c \propto 1/L$. In fact, we found that this
$1/L$ dependence is quite robust as it holds whenever there
is a 1-step RSB solution. In a later section we will
analyze the physical consequences of this behavior
and show that it can be simply explained via the path integral
mapping to the Schr\"odinger equation.

\begin{figure}
\centerline{\epsfysize 8.5cm \epsfbox{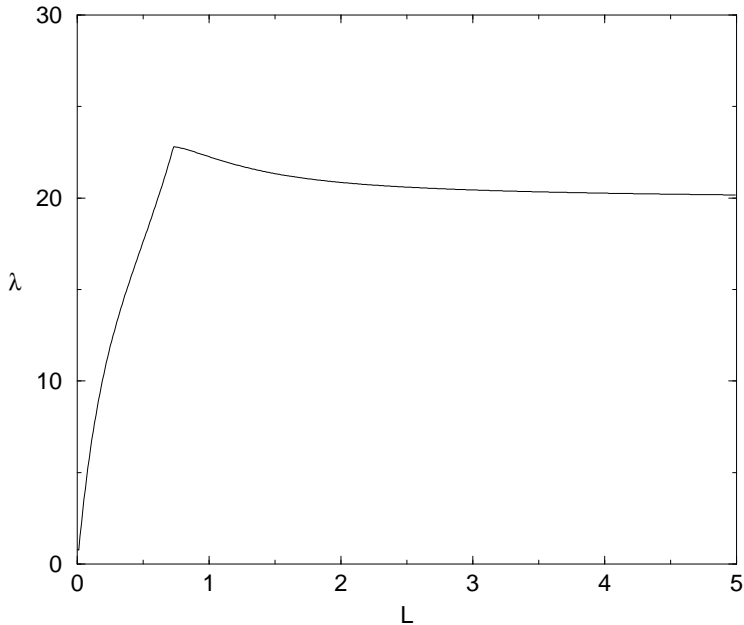}} \vspace{3mm}
\caption{Plot of $\lambda$ vs. $L$. The parameters are the same 
as those used in Fig.~\ref{rf1}.
}
\label{lplot}
\end{figure}

In Fig.~\ref{lplot} we plot the parameter $\lambda$ vs. $L$.
The discontinuity is of course at $L_c$, after which we plot
only the 1-step RSB solution.  Notice that for $L>L_c$ $\lambda$ is 
essentially constant, which by Eq.~(\ref{rsqf}) implies that $\AVBR$ is also
constant.
This result is consistent with the approximate analytic solution in 
 Ref.~\cite{gold}, where it was found that 
\begin{equation}
\lambda=\frac{d^{4/(4-d)}}{(2\pi)^{2d/(4-d)}}
(\beta^2 M)^{(4+d)/(4-d)}(g|\log \mu |)^{4/(4-d)},
\label{aequation}
\end{equation}
which is a constant independent of $L$.

\section{Physical interpretation of the 1-step RSB solution}

In this section we study the physical interpretation of
the replica symmetry breaking solution.  Our purpose is
to see if the underlying physical picture predicted
by 1-step RSB  is indeed consistent with the presence of exponentially
localized eigenstates. The following analysis is valid for
a very long polymer (large $L$) when the system becomes glassy.
We begin by evaluating the replicated partition sum defined as
\begin{equation}
\tilde{Z}_n(\{\BR_a\})=\int_{\BR_a(0)=\BR_a}^{\BR_a(L)=\BR_a}
\prod_{a=1}^n[d\BR_a]\exp(-\beta h_n),
\label{Zrep}
\end{equation}
where $h_n$ is the quadratic trial Hamiltonian in Eq.~(\ref{trialh}).
Since $h_n$ is quadratic the path integrals can be evaluated 
analytically and the final result can we written in the form
\begin{equation}
\tilde{Z}_n(\{\BR_a\})={\rm const.} \times \exp \left(-\frac{1}{2}
\sum_{ab} Q_{ab}^{-1} \BR_a \cdot \BR_b \right).
\label{ZQ}
\end{equation} 
The details of this calculation along with the relationship
between the matrices $Q_{ab}$ and $p_{ab}$ are given in the Appendix.
Now, since $p_{ab}$ was parametrized according to the 1-step RSB scheme,
it implies that $Q_{ab}$ can also be parametrized in the same way.

Mezard and Parisi \cite{MP} discuss the interpretation of a
representation of the form (\ref{ZQ}) for the case of directed
polymers. In particular they show how to deduce
the structure of the probability distribution 
\begin{eqnarray}
P_V(\BR)=\tilde{Z}_V (\BR,L)/\int d \BR \tilde{Z}_V(\BR,L),
\label{PV}
\end{eqnarray}
which is the probability of finding a 
polymer loop that passes through $\BR$ for a given realization
(which we denote by $V$) of the the random potential. Here $\tilde{Z}_V
(\BR,L)$ is just the partition sum $Z(\BR,\BR;L)$ as given in
Eq.~(\ref{Z}). This probability is related to the replicated partition
function given in Eq.~(\ref{Zrep}) by
\begin{eqnarray}
P_V(\BR)= \lim_{ n \rightarrow 0} \int d\BR_2 \cdots d\BR_n\
\tilde{Z}_n(\{\BR_a\})_{\BR_1=\BR}. 
\label{PVrel}
\end{eqnarray}
Mezard and Parisi's analysis has to be adapted for the case of real
(non-directed) polymers of length $L$ in a random potential which 
is independent of time. The changes will be pointed out below.
 
If  $Q_{ab}$ is parametrized by $\{\tilde{q},q(x)\}$ such that
\begin{eqnarray}
q(x)= \left\{ \begin{array}{l}
q_0 \ \ \ x < x_c \  \\
q_1 \ \ \ x > x_c \ ,
\end{array}
\right.
\end{eqnarray}
one proceeds to obtain $P_V(\BR)$ by the following procedure: 

\begin{flushleft} 1. For each sample ( a realization of the random potential)
generate a random variable $\BR_0$ which is picked from the distribution
\begin{equation}
{\cal P}(\BR_0)=\frac{1}{(2\pi q_0)^{d/2}}
\exp  \left( -\frac{\BR_0^2}{2q_0}\right).
\end{equation}

2. Consider a set of ``states'' labeled by the index
$\alpha$ whose physical meaning will be elucidated shortly. Each of these
states is characterized by a weight $W_\alpha$ and a position variable
$\BR_\alpha$. Given $\BR_0$, the variables $\BR_\alpha$ are an infinite
set of uncorrelated random variables distributed according to
\begin{equation}
{\cal P}(\BR_1,\BR_2,\cdots)=\prod_\alpha\frac{1}{(2\pi(q_1-q_0))^{d/2}}
\exp  \left( -\frac{(\BR_{\alpha}-\BR_0)^2}{2(q_1-q_0)}    \right).
\label{positiondist}
\end{equation} 
The distribution of weights will be discussed below. 

3. Given these ``states'' for a given sample, The probability
distribution $P_V(\BR)$ for that sample has the form
\begin{equation}
P_V(\BR)=\sum_{\alpha} W_{\alpha}
\frac{1}{(2\pi(\tilde{q}-q_1))^{d/2}}
 \exp \left( 
-\frac{(\BR-\BR_{\alpha})^2}{2(\tilde{q}-q_1)} \right).
\label{PVR}
\end{equation}
\end{flushleft}

The weights $W_\alpha$ are given in terms of some ``free energy'' variables
$f_\alpha$:
\begin{eqnarray}
W_\alpha = \frac{\exp(-\beta f_\alpha)}{\sum_{\gamma} \exp(-\beta f_{\gamma}) }.
\label{falpha}
\end{eqnarray}
These free energy variables are chosen from an exponential
distribution
\begin{eqnarray}
P[f_\alpha] \propto \exp (x_c \beta f_\alpha) \ \theta(f-\bar{f}),
\label{distf}
\end{eqnarray}
where $\bar{f}$ is an upper cutoff.

What is the meaning of these variables in the present case?
To determine the weights $W_{\alpha}$ we compare Eq.~(\ref{PVR}) to the
eigenfunction expansion given in Eq.~(\ref{loopexp}). From
Eq.~(\ref{loopexp}) together with Eq.~(\ref{PV}) it becomes clear that 
\begin{equation}
P_V(\BR)=\sum_\alpha A_\alpha |\Phi_\alpha(\BR)|^2,
\end{equation}
where 
\begin{equation}
A_\alpha = \frac{\exp(-\beta L E_\alpha)}{\sum_{\gamma} \exp(-\beta L 
E_{\gamma}) }.
\label{eigenweights}
\end{equation}
Comparing Eq.~(\ref{PVR}) and Eq.~(\ref{eigenweights}) it becomes
obvious that 
\begin{eqnarray}
W_\alpha=A_\alpha,
\label{WA}
\end{eqnarray}
and 
\begin{equation}
\Phi_\alpha^2(\BR) \propto  \exp \left( 
-\frac{(\BR-\BR_{\alpha})^2}{2(\tilde{q}-q_1)} \right).
\end{equation}
Hence, the ``states'' labeled by $\alpha$ are in our case the actual eigenstates
of the imaginary time Schr\"odinger equation (\ref{eigenvalue}). These
are localized tail states centered at position $\BR_{\alpha}$ with an 
associated ``weight'' $W_\alpha$.
Thus the 1-step RSB solution approximates the tail states by a fixed
Gaussian form.
The width of these Gaussians ($w_{0}$) as defined by 
Eq.~(\ref{widtheq}) satisfies $w_0^2 = d(\tilde{q}-q_1)$,
which for large $L$ can be shown to converge to 
$w_0^2 \sim d/(2\beta \sqrt{\lambda M)}$ (see Appendix). Since 
$\lambda$ becomes constant for large $L$  then so does the width
$w_0$.  Thus the 1-step RSB solution approximates all
the localized states by a Gaussian of constant width $w_0$, which is
the typical size of a Gaussian chain embedded in a random potential.
It also becomes evident that the free energies $f_\alpha$ are equal
to $ L E_\alpha$. This make sense if we think of $|E_\alpha|$ as
representing the binding energy per monomer, and thus $f_\alpha= L E_\alpha$
represent the total energy of the chain. Alternatively one can start 
with the quantum particle picture and identify $f_\alpha$ with
$E_\alpha$ but then in the transformation from a quantum particle to a
polymer one has to replace $\beta$ by $\beta L $, and thus one ends up
with the same expressions.

These arguments lead us to expect that within the 1-step RSB scheme,
the energy variables $E_\alpha$ are independent random variables
taken from an exponential distribution: 
\begin{equation}
P[E_\alpha] \propto e^{\beta L x_c E_\alpha} \theta(\overline{E}-E_\alpha) 
\ \ ,
\label{eigendist2}
\end{equation}
with $\overline{E}$ being some energy scale determined by the upper
cutoff of the tail region.
We will now argue that the distribution given above is just
the expected distribution of ground-state energies i.e. the probability
of finding the lowest energy level to have energy $E$. We first review some
very basic results of extreme value statistics as presented in
Ref.~\cite{BM}.  Given $K$ independent and identically distributed
random variables $E_i$, pulled from a distribution of the form
\begin{equation}
\tilde{P}(E)=\frac{A}{|E|^\alpha} \exp(-B|E|^\delta),
\label{eigendist}
\end{equation}
the probability that the lowest of the $K$ energies is $E$ (for
$E \rightarrow -\infty $ and $K \rightarrow \infty$) is
given by 
\begin{equation}
P(E) \propto \exp{[B\delta |E_c|^{\delta-1} E]} \ ,
\label{extreme}
\end{equation}
where 
\begin{equation}
E_c=-\left( \frac{\log(K)}{B} \right)^{1/\delta}.
\label{Ec}
\end{equation}
The value of $E_c$, the lowest energy expected to be attained in K
trials, is easily obtained from
\begin{eqnarray}
\int_{-\infty}^{E_c} dE \tilde{P}(E) \simeq 1/K.
\label{Ecform}
\end{eqnarray}

The reason why we chose a distribution of the form given
in Eq.~(\ref{eigendist}) is that in $d=1$ the probability $\tilde{P}(E)$ is 
known to have that form exactly for the case
of delta correlated random potentials (see
\cite{halperin}).  For $d>1$ Lifshits \cite {lifshits} argued that 
the form given by Eq.~(\ref{eigendist}) is also valid.
Our goal now is to see if the distribution
Eq.~(\ref{eigendist2}) derived using the 1-step RSB solution is
indeed consistent with the distribution Eq.~(\ref{extreme}) 
predicted using extreme value statistics.
 
Comparing Eq.~(\ref{eigendist2}) and Eq.~(\ref{extreme}) we find
that for consistency the break point should satisfy
\begin{equation}
x_c=\frac{\delta}{\beta L} B^{1/\delta}(\log(K))^{(\delta-1)/\delta}.
\end{equation}
Notice that the $1/L$ behavior of $x_c$ is exactly the same
as was found analytically for large $L$ in Ref. \cite{gold}
and numerically for any $L>L_c$ in the present work.  

We can go further by using the fact that the number of energy levels  
$K$, within a fixed
energy interval is directly proportional to the system size, which in
our formulation is effectively determined by $\mu$.  
Assuming $\log(K) \propto |\log(\mu)|$ and comparing to
the approximate solution for $x_c$ in Eq.~(\ref{zequation})
we find that $\delta=(4-d)/2$ and $B \propto 1/g$.  
Now $\tilde{P}(E)$ is just proportional to the density of states
$\rho(E)$, and it is known exactly in one dimension. Indeed, when $d=1$,
$\delta=3/2$ and $B \propto 1/g$. For $2 \leq d < 4$, $\delta$ agrees with the
result derived by Lifshits \cite{lifshits}.
Hence, the exponent $\delta$ and the disorder dependence of
$B$ is correctly predicted by the 1-step RSB solution.
The exponent $\alpha$ cannot be found since the
statistics of the lowest energy depends only on the exponential
tail.  

The 1-step RSB solution is thus in agreement with 
Eq.~(\ref{extreme}).
Since $x_c \propto 1/L$ then the distribution of the energies
predicted from Eq.~(\ref{eigendist2}) is independent of $L$. 
Note also that the value of $E_c$ which gives a typical
lowest energy for a given volume of the system coincides with the
expected value given in Eq.~(\ref{depth}) when one substitutes for
$B\sim 1/g$, $\delta=(4-d)/2$ and $\ln K \sim \ln {\cal V}$ in
Eq.~(\ref{Ec}).  The localization length which is the width of
a localized state is known to be related to the energy of the state by
\begin{eqnarray}
\ell \sim \frac{1}{\beta |2 M E|^{1/2}}.
\label{lc}
\end{eqnarray}
If we use $E_c$ for the value of the lowest energy in a volume ${\cal
V}$ we get an estimate for $R_F$, the radius of gyration of a free
chain which is in agreement with Eq.~(\ref{lnV}). 
 
When $L \rightarrow \infty$ most of the weights $W_{\alpha} 
\rightarrow 0$ and only
the weight with the lowest $E_{\alpha}$ will contribute
significantly to the distribution $P_V(\BR)$.  So in this limit $P_V(\BR)$
will be approximated by a single Gaussian located somewhere
in the sample.  
This result is consistent with ground-state dominance as predicted
by the corresponding eigenfunction expansion.  

We now consider the distribution $P(\BR_\alpha)$ given in 
Eq.~(\ref{positiondist}). This is just the distribution for
the localization centers $\BR_\alpha$ for a given value of $\BR_0$.
Hence, we can calculate
the average distance between the localized states for a given sample.
We find that the width $w$ of the Gaussian $P(\BR_\alpha)$ 
satisfies $w^2=d(q_1-q_0)$. For small $\mu$ and large $L$ we 
find (see Appendix) that $w^2 \approx d/(\beta \mu Lx_c)$.  Using 
the analytic approximation for $x_c$ we find that 
\begin{equation}
w^2 \sim \frac{1}{\mu} 
g^{2/(4-d)}|\log(\mu)|^{(d-2)/(4-d)}.
\end{equation}
We can also find the average distance from the origin to a
localization center $\BR_\alpha$.  To do this we have to compute
the average probability of find a localization center at $\BR_\alpha$.
This is
\begin{eqnarray}
\overline{P}(\BR_\alpha)\propto \int d\BR_0 \exp\left(-\BR_0^2/(2q_0)\right)
\exp\left(-(\BR_\alpha-\BR_0)^2/(2(q_1-q_0)\right) \nonumber \\
= \exp\left(-\BR_\alpha^2/(2q_1)\right).
\end{eqnarray}
If we estimate the average distance to a localization center to
be the width of the Gaussian $\overline{P}(\BR_\alpha)$, then
$w^2=dq_1\approx d/(\beta\mu L x_c)$ which is the same as the average
distance $w$ between tail states derived above.  This is because the 
fluctuations of $\BR_0$ are small compared to the fluctuations of 
$\BR_\alpha-\BR_0$ when the volume is large. Notice that this result is 
consistent with our calculation for $\ABRQ$ in Eq.~(\ref{shift}),
since for $L$ large $\ABRQ \sim  d/(\beta \mu Lx_c)$.
These quantities should be close since both give an approximation 
to the average distance to the ground state localization center.

So far we have studied the physical interpretation of the
1-step RSB solution, but recall that for $L < L_c$ there was only a RS 
solution.  It is interesting to study the physical implications of this
RS solution.  In this case the probability distribution
$P_V(\BR)$ will have the simple form
\begin{equation}
P_V(\BR)=
\frac{1}{(2\pi(\tilde{q}-q_0))^{d/2}}
 \exp \left( 
-\frac{(\BR-\BR_{0})^2}{2(\tilde{q}-q_0)} \right),
\end{equation}
where the variable $\BR_0$ is taken from the distribution
\begin{equation}
{\cal P}(\BR_0)=
\frac{1}{(2\pi q_0)^{d/2}}
 \exp \left( 
-\frac{\BR_{0}^2}{2q_0} \right).
\label{rsshift}
\end{equation}
Notice that for a given sample $P_V(\BR)$ is approximated by a Gaussian
shifted from the origin by a distance $\BR_0$. 
The width of this Gaussian can be shown to be essentially independent
of disorder in the limit $\mu \rightarrow 0$, as the width $w_0$ satisfies
$w_0^2 \sim d(\beta \mu L)^{-1}$,
which is the result for the case of zero disorder.    The average
shift from the origin can be estimated by the width $w_1$ of the Gaussian
in Eq.~(\ref{rsshift}), which satisfies $w_1^2 \sim g\beta^{(d+2)/2}
L^{(d+2)/2} \mu^{(d-2)/2}$, for the case of delta correlations
and large volume ($\mu \rightarrow 0)$. This result is similar
to that found for the case of long range correlated random media
in which it was found that the only effect of disorder is to shift
the center of mass of the polymer chain from the origin.

We can conclude that the physical interpretation of the
1-step RSB solution is consistent with that given by the
eigenfunction expansion.  Clearly, the 1-step RSB solution 
captures the localized states and also correctly predicts
some important features of the eigenvalue distribution.
However, there are differences and these reveal the limitations
of the 1-step RSB solution. For example, all the localized states are
approximated by the same Gaussian profile when in fact the
localization lengths should increase with energy.  

\section{Concluding Remarks}

The findings in this paper allow us to give a fairly complete
description of a polymer in random medium.  We find that the 
size of a short polymer chain behaves as if there is no disorder,
but that the position of the center of mass of the chain is 
strongly affected by the random media.  Applying the replica variational
method we find that short chains are well described by a RS
solution.  Also, using an eigenfunction expansion
we show that the partition sum of short chains is dominated by
extended eigenstates.  For long chains the physical picture becomes
more interesting and complex.  We find that long chains are 
likely to be localized in regions of the sample where there
is very low potential energy.  These regions correspond to localized
tail states of the corresponding Schr\"odinger equation.  As the length
of the chain increases the number of dominant conformations decreases
until finally the chain is essentially localized in one small
region, consistent with ground state dominance. In terms of the
mapping from a chain to a quantum particle, a big chain length
corresponds to a very low temperature for the quantum particle.  

We show that the onset of the variational
RSB solution to the stationarity equations corresponds to the 
dominance of localized tail states. This 1-step RSB solution
correctly describes the glassy characteristics of the polymer
chain such as large sample-to-sample variations. We have demonstrated
the clear physical picture associated with the abstract RSB solution. 

Our analysis also suggests a direct connection between the size of
a polymer in a random potential with short range correlations, and the
localization length associated with the lowest energy state in a
system of finite volume ${\cal V}$. Given the density of states $\rho(E)$ we can estimate the
lowest energy $E_c$ by 
\begin{eqnarray}
\int_{-\infty}^{E_c} dE \rho(E) \simeq 1/{\cal V}.
\label{lowest}
\end{eqnarray}
The size of a free Gaussian chain, in the long chain limit, is then given by
\begin{eqnarray}
R_F \sim |E_c|^{-1/2}.
\label{rfe}
\end{eqnarray}
It is remarkable that starting from an annealed average (the density
of states involves the average of the partition function) and using
extreme value statistics plus the known relationship between the
localization length and the value of the tail state energy, we can
predict the behavior of a quenched average like the chain size.

It is well known that glassy systems are notoriously difficult 
to simulate using the Monte Carlo method since 
equilibration times are exceedingly large. In the context of a 
polymer chain in a random medium, we expect that a 
long chain will get trapped in the deep wells of the
random potential and any local updating procedure, such as the 
Metropolis scheme, will not be able to find the true minimum 
energy configurations in a reasonable time. A workaround is to use a
simulated annealing procedure \cite{chen}.
Of course, these traps are just those regions of the sample where an 
eigenstate would be localized. They correspond to localized states 
with energies above the true ground state that exists for a
finite-size system. Since the partition sum 
for a long chain is dominated by localized tail states
those chain configurations that are in the vicinity
of such a state are overwhelmingly more favorable than
other configurations. All the classic characteristics of a glassy
system emerge (like trapping, metastability, aging),
and they have a clear physical origin in terms of the
localization picture.
\begin{acknowledgments}
This research is supported by the US Department of Energy (DOE), grant No.
DE-G02-98ER45686. Y. Y. G. thanks the Weizmann Institute for a Meyerhoff
Visiting Professorship during which some of this reseach has been done.
\end{acknowledgments}

\newpage  
\appendix
\section*{}
\newcommand{\BV}{{\bf{V}}}
In this appendix we derive the relation between the 
matrix $p_{ab}$ (see Eq.~(\ref{pmat}) ) characterizing the variational
hamiltonian and the matrix $Q_{ab}$ defined in Eq.~(\ref{ZQ}).  We
start from
\begin{eqnarray}
Z_n(\{{\mathbf{R}}'_a\},\{{\mathbf{R}}_a\};L)=
 \int_{{\mathbf{R}}_a(0)={\mathbf{R}}'_a}^{{\mathbf{R}}_a(L)={\mathbf{R}}_a} 
\prod_{a=1}^n[d{\mathbf{R}}_a(u)]\exp(-\beta h_n),
\label{A1}
\end{eqnarray} 
with
\begin{eqnarray}
h_n=\frac{1}{2}\int_0^L du \sum_{a} \left[ M\left(\frac{d {\mathbf{R}}_a (u)}{d 
u}\right)^2 +\lambda {\mathbf{R}}^2_a (u) \right] \nonumber \\
+\frac{1}{2L}\int^L_0 du 
\int^L_0 du' \sum_{ab} p_{ab} \BR_a (u) \cdot \BR_b (u').
\label{A2}
\end{eqnarray}
Let $O_{ab}$ be an orthogonal matrix which diagonalizes $p_{ab}$ by a
similarity transformation, and let $p_c$, $c=1,\cdots,n$ be the
eigenvalues of $p_{ab}$. We can express $Z_n$ in the form
\begin{eqnarray}
Z_n=\int \prod_c d\BL_c \exp\left(-\sum_c\frac{\BL_c}{2}\right)
\int_{{\mathbf{R}}_a(0)={\mathbf{R}}'_a}^{{\mathbf{R}}_a(L)={\mathbf{R}}_a} 
\prod_{a=1}^n[d{\mathbf{R}}_a(u)]\exp(-\beta h_n(\{\BL_c\})),
\label{A3}
\end{eqnarray}
with
\begin{eqnarray}
h_n(\{\BL_c\})=\frac{1}{2}\int_0^L du \sum_{a} \left[
M\left(\frac{d {\mathbf{R}}_a (u)}{d 
u}\right)^2 +\lambda {\mathbf{R}}^2_a (u) + \sum_c 
\sqrt{-p_c/(\beta L)} O_{ac} \ \BL_c \cdot {\mathbf{R}}_a(u).
\right]
\label{A4}
\end{eqnarray}
The path integral can now be carried out using Eqns.~(3.39-3.41) in
\cite{feynman}. The result is 
\begin{eqnarray}
Z_n=\int \prod_c d\BL_c \exp\left(-\sum_c\frac{\BL_c}{2}\right)
e^{-\beta \Phi},
\label{A5}
\end{eqnarray}
where
\begin{eqnarray}
\Phi &=& A\left(\sum \BR_a^2+\sum\BR_a^{\prime 2}\right)+B\left(\sum 
\BR_a+\sum \BR'_a\right)\cdot \BV_a 
+2C\sum \BR_a\cdot \BR'_a+ D\sum \BV_a^2,
\label{phi}
\end{eqnarray}
with
\begin{eqnarray}
A &=& \frac{1}{2} \sqrt{\lambda M} \coth \left( L\sqrt{\lambda/M}
\right)  \\
B &=& \sqrt{\lambda M}
\left[\cosh\left( L\sqrt{\lambda/M }  \right)-1   \right]
\left[\sinh \left( L\sqrt{\lambda/M}\right)\right]^{-1}  
 \\
C &=& -\frac{1}{2} \sqrt{\lambda M}\left[(\sinh \left( L\sqrt{\lambda/M}
\right)\right]^{-1}  \\
D &=& B-L\lambda/2, \\
\BV_a &=& \frac{1}{\lambda}\sum_c\sqrt{-p_c/(\beta L)} O_{ac} \ \BL_c .
\label{A7}
\end{eqnarray}
Notice that 
\begin{eqnarray}
\sum_a \BV_a^2=-\frac{1}{\lambda^2 \beta L}\sum_a p_a \BL_a^2
\label{v2}
\end{eqnarray}

We now put $\BR_a=\BR'_a$ and integrate out over $\BL_c$ to obtain
\begin{eqnarray}
Z_n(\{\BR_a\},\{\BR_a\};L)={\cal N} \exp\left(-2 \beta\sum_{ab}   [
(A+C)\delta_{ab}-\beta B^2 {\cal M}_{ab}]\BR_a\cdot \BR_b\right),
\label{A9}
\end{eqnarray}
with
\begin{eqnarray}
{\cal M}=\frac{1}{2 \beta D}\left({\mathbf{1}}-\frac{\lambda^2 L}{2
D}{\mathbf{p}}^{-1}\right)^{-1}.
\label{A10}
\end{eqnarray}
Here we denoted by  ${\mathbf{1}}$ the unit $n\times n$ matrix, and
${\mathbf{p}}$ stands for the matrix $p_{ab}$. Comparing with
Eq.~(\ref{ZQ}) we find that
\begin{eqnarray}
Q_{ab}=\frac{1}{4 \beta} \{[(A+C){\mathbf{1}}-\beta B^2 
{\cal M}]^{-1}\}_{ab}.
\label{Qf}
\end{eqnarray}

In the Appendix of Ref.~\cite{MP} one can find a formula for the
inverse of a hierarchical matrix. For the one-step RSB case, 
the inverse of a matrix
${\bf p}=\{\tilde{p},p_0,p_1\}$ is given by ${\bf q}={\bf p}^{-1}$ where
\begin{eqnarray}
\tilde{q}&=&\frac{1}{(\tilde{p} - p_0 x_c - p_1 (1 - x_c))}\left(1 - 
        \frac{p_0}{\tilde{p} - p_0 x_c - p_1 (1 - x_c)} - \frac{(p_1 - 
              p_0) (1 - x_c)}{\tilde{p} - p_1}\right), \\
q_0&=& -\frac{p_0}{(\tilde{p} - p_0 x_c - 
          p_1 (1 - x_c))^2}, \\ 
q_1&=& -\frac{1}{\tilde{p} - p_0 x_c - p_1 (1 - x_c)} \left(
\frac{p_1 - p_0}{\tilde{p} - p_1} +
         \frac{p_0}{\tilde{p} - p_0 x_c - p_1 (1 - x_c)}\right).
\label{inverse}
\end{eqnarray}

In the limit of large $L$ we have\begin{eqnarray}
A \rightarrow \frac{1}{2} \sqrt{\lambda M} \\
B \rightarrow \sqrt{\lambda M} \\
C \rightarrow 0 \\
D \rightarrow -\frac{L \lambda}{2} + \sqrt{\lambda M}.
\label{lim}
\end{eqnarray}
In addition we use the one step breaking results found in
Ref.~\cite{gold}. Evaluating the matrix $Q$ in the limit of small
$\mu$ we find to leading order
\begin{eqnarray}
\tilde{q} &=& \frac{1}{Lx_c\beta \mu} +
\frac{s_0}{L\beta\mu^2}+\frac{1}{2\beta \sqrt{M\lambda}}+\cdots \\
q_0&=&\frac{s_0}{L\beta\mu^2}+\cdots \\
q_1&=&\frac{1}{Lx_c\beta \mu} +
\frac{s_0}{L\beta\mu^2}+\cdots
\label{approxQ}
\end{eqnarray}
with the parameters $\lambda$, and $x_c$ given in Eq.~(\ref{lamfin}) and 
(\ref{xcf}) respectively and $s_0$ is given approximately by
\cite{gold}
\begin{eqnarray}
s_{0}  ={\mathrm const}.\times g^{(2-d)/4}\beta L(\beta\sqrt{M\lambda
})^{-d(d+2)/8}\mu^{d/2+1}\left|  \ln\mu\right|  ^{(d+2)/4}.
\label{s0f}
\end{eqnarray}
We have also used the fact that $\Sigma=\lambda$ in this limit.

\end{document}